\documentclass[twocolumn,tighten,twocolappendix]{aastex63}
\pdfoutput=1 
\usepackage{amsmath,amstext}
\usepackage[T1]{fontenc}
\usepackage{ae,aecompl}
\usepackage[utf8]{inputenc}
\usepackage{apjfonts} 
\usepackage[figure,figure*]{hypcap}
\usepackage{enumitem}
\usepackage{bm}
\usepackage{graphicx}
\input{hyperlink-year-only-natbib-patch}

\reportnum{DES-2019-0491}
\reportnum{FERMILAB-PUB-19-606-AE}

\newcommand*{\mailto}[1]{\href{mailto:#1}{#1}}
\newcommand*{\http}[1]{\href{http://#1}{#1}}
\newcommand*{\https}[1]{\href{https://#1}{#1}}

\newcommand{\K}{\ensuremath{{\rm K}}}
\newcommand{\s}{\ensuremath{{\rm s}}}

\newcommand{\rvir}{\ensuremath{{R_{\rm vir}}}}

\shorttitle{Milky Way Satellite Census. II.}
\shortauthors{Nadler \& Wechsler et al.}
\defcitealias{PaperI}{Paper I}

\begin{document}

\title{Milky Way Satellite Census. II. Galaxy--Halo Connection Constraints Including the Impact of the Large Magellanic Cloud}

\author[0000-0002-1182-3825]{E.~O.~Nadler}
\affiliation{Department of Physics, Stanford University, 382 Via Pueblo Mall, Stanford, CA 94305, USA; \mailto{enadler@stanford.edu}}
\affiliation{Kavli Institute for Particle Astrophysics \& Cosmology, P. O. Box 2450, Stanford University, Stanford, CA 94305, USA; \mailto{rwechsler@stanford.edu}}
\author[0000-0003-2229-011X]{R.~H.~Wechsler}
\affiliation{Department of Physics, Stanford University, 382 Via Pueblo Mall, Stanford, CA 94305, USA}
\affiliation{Kavli Institute for Particle Astrophysics \& Cosmology, P. O. Box 2450, Stanford University, Stanford, CA 94305, USA}
\affiliation{SLAC National Accelerator Laboratory, Menlo Park, CA 94025, USA}
\author[0000-0001-8156-0429]{K.~Bechtol}
\affiliation{Physics Department, 2320 Chamberlin Hall, University of Wisconsin--Madison, 1150 University Avenue, Madison, WI 53706-1390, USA}
\affiliation{LSST, 933 North Cherry Avenue, Tucson, AZ 85721, USA}
\author[0000-0002-1200-0820]{Y.-Y.~Mao}
\altaffiliation{NHFP Einstein Fellow}
\affiliation{Department of Physics and Astronomy, Rutgers, The State University of New Jersey, Piscataway, NJ 08854, USA}
\affiliation{Department of Physics and Astronomy, University of Pittsburgh, Pittsburgh, PA 15260, USA}
\affiliation{Pittsburgh Particle Physics, Astrophysics and Cosmology Center (PITT PACC), University of Pittsburgh, Pittsburgh, PA 15260, USA}
\author[0000-0001-5417-2260]{G.~Green}
\affiliation{Max Planck Institute for Astronomy, K\"onigstuhl 17 D-69117, Heidelberg, Germany}
\author[0000-0001-8251-933X]{A.~Drlica-Wagner}
\affiliation{Fermi National Accelerator Laboratory, P. O. Box 500, Batavia, IL 60510, USA}
\affiliation{Kavli Institute for Cosmological Physics, University of Chicago, Chicago, IL 60637, USA}
\affiliation{Department of Astronomy and Astrophysics, University of Chicago, Chicago, IL 60637, USA}
\author[0000-0001-5435-7820]{M.~McNanna}
\affiliation{Physics Department, 2320 Chamberlin Hall, University of Wisconsin--Madison, 1150 University Avenue, Madison, WI 53706-1390, USA}
\author[0000-0003-3519-4004]{S.~Mau}
\affiliation{Kavli Institute for Cosmological Physics, University of Chicago, Chicago, IL 60637, USA}
\affiliation{Department of Astronomy and Astrophysics, University of Chicago, Chicago, IL 60637, USA}
\author[0000-0002-6021-8760]{A.~B.~Pace}
\affiliation{Department of Physics, Carnegie Mellon University, Pittsburgh, PA 15312, USA}
\affiliation{George P. and Cynthia Woods Mitchell Institute for Fundamental Physics and Astronomy, and Department of Physics and Astronomy, Texas A\&M University, College Station, TX 77843,  USA}
\author{J.~D.~Simon}
\affiliation{Observatories of the Carnegie Institution for Science, 813 Santa Barbara St., Pasadena, CA 91101, USA}
\author[0000-0003-4307-634X]{A.~Kravtsov}
\affiliation{Kavli Institute for Cosmological Physics, University of Chicago, Chicago, IL 60637, USA}
\author[0000-0002-8446-3859]{S.~Dodelson}
\affiliation{Department of Physics, Carnegie Mellon University, Pittsburgh, Pennsylvania 15312, USA}
\author{T.~S.~Li}
\altaffiliation{NHFP Einstein Fellow}
\affiliation{Observatories of the Carnegie Institution for Science, 813 Santa Barbara Street, Pasadena, CA 91101, USA}
\affiliation{Department of Astrophysical Sciences, Princeton University, Peyton Hall, Princeton, NJ 08544, USA}
\affiliation{Fermi National Accelerator Laboratory, P. O. Box 500, Batavia, IL 60510, USA}
\affiliation{Kavli Institute for Cosmological Physics, University of Chicago, Chicago, IL 60637, USA}
\author[0000-0001-5805-5766]{A.~H.~Riley}
\affiliation{George P. and Cynthia Woods Mitchell Institute for Fundamental Physics and Astronomy, and Department of Physics and Astronomy, Texas A\&M University, College Station, TX 77843,  USA}
\author[0000-0002-8226-6237]{M.~Y.~Wang}
\affiliation{Department of Physics, Carnegie Mellon University, Pittsburgh, Pennsylvania 15312, USA}
\author{T.~M.~C.~Abbott}
\affiliation{Cerro Tololo Inter-American Observatory, National Optical Astronomy Observatory, Casilla 603, La Serena, Chile}
\author{M.~Aguena}
\affiliation{Laborat\'orio Interinstitucional de e-Astronomia - LIneA, Rua Gal. Jos\'e Cristino 77, Rio de Janeiro, RJ---20921-400, Brazil}
\affiliation{Departamento de F\'isica Matem\'atica, Instituto de F\'isica, Universidade de S\~ao Paulo, CP 66318, S\~ao Paulo, SP, 05314-970, Brazil}
\author[0000-0002-7069-7857]{S.~Allam}
\affiliation{Fermi National Accelerator Laboratory, P. O. Box 500, Batavia, IL 60510, USA}
\author[0000-0002-0609-3987]{J.~Annis}
\affiliation{Fermi National Accelerator Laboratory, P. O. Box 500, Batavia, IL 60510, USA}
\author{S.~Avila}
\affiliation{Instituto de Fisica Teorica UAM/CSIC, Universidad Autonoma de Madrid, E-28049 Madrid, Spain}
\author{G.~M.~Bernstein}
\affiliation{Department of Physics and Astronomy, University of Pennsylvania, Philadelphia, PA 19104, USA}
\author{E.~Bertin}
\affiliation{CNRS, UMR 7095, Institut d'Astrophysique de Paris, F-75014, Paris, France}
\affiliation{Sorbonne Universit\'es, UPMC Univ Paris 06, UMR 7095, Institut d'Astrophysique de Paris, F-75014, Paris, France}
\author{D.~Brooks}
\affiliation{Department of Physics \& Astronomy, University College London, Gower Street, London, WC1E 6BT, UK}
\author{D.~L.~Burke}
\affiliation{SLAC National Accelerator Laboratory, Menlo Park, CA 94025, USA}
\affiliation{Kavli Institute for Particle Astrophysics \& Cosmology, P. O. Box 2450, Stanford University, Stanford, CA 94305, USA}
\author[0000-0003-3044-5150]{A.~Carnero~Rosell}
\affiliation{Centro de Investigaciones Energ\'eticas, Medioambientales y Tecnol\'ogicas (CIEMAT), Madrid, Spain}
\affiliation{Laborat\'orio Interinstitucional de e-Astronomia - LIneA, Rua Gal. Jos\'e Cristino 77, Rio de Janeiro, RJ - 20921-400, Brazil}
\author[0000-0002-4802-3194]{M.~Carrasco~Kind}
\affiliation{National Center for Supercomputing Applications, 1205 West Clark Street, Urbana, IL 61801, USA}
\affiliation{Department of Astronomy, University of Illinois at Urbana-Champaign, 1002 West Green Street, Urbana, IL 61801, USA}
\author[0000-0002-3130-0204]{J.~Carretero}
\affiliation{Institut de F\'{\i}sica d'Altes Energies (IFAE), The Barcelona Institute of Science and Technology, Campus UAB, E-08193 Bellaterra (Barcelona), Spain}
\author{M.~Costanzi}
\affiliation{INAF--Osservatorio Astronomico di Trieste, via G. B. Tiepolo 11, I-34143 Trieste, Italy}
\affiliation{Institute for Fundamental Physics of the Universe, Via Beirut 2, I-34014 Trieste, Italy}
\author{L.~N.~da Costa}
\affiliation{Laborat\'orio Interinstitucional de e-Astronomia - LIneA, Rua Gal. Jos\'e Cristino 77, Rio de Janeiro, RJ - 20921-400, Brazil}
\affiliation{Observat\'orio Nacional, Rua Gal. Jos\'e Cristino 77, Rio de Janeiro, RJ - 20921-400, Brazil}
\author[0000-0001-8318-6813]{J.~De~Vicente}
\affiliation{Centro de Investigaciones Energ\'eticas, Medioambientales y Tecnol\'ogicas (CIEMAT), Madrid, Spain}
\author[0000-0002-0466-3288]{S.~Desai}
\affiliation{Department of Physics, IIT Hyderabad, Kandi, Telangana 502285, India}
\author[0000-0002-4876-956X]{A.~E.~Evrard}
\affiliation{Department of Astronomy, University of Michigan, Ann Arbor, MI 48109, USA}
\affiliation{Department of Physics, University of Michigan, Ann Arbor, MI 48109, USA}
\author{B.~Flaugher}
\affiliation{Fermi National Accelerator Laboratory, P. O. Box 500, Batavia, IL 60510, USA}
\author{P.~Fosalba}
\affiliation{Institute of Space Sciences (ICE, CSIC),  Campus UAB, Carrer de Can Magrans, s/n, E-08193 Barcelona, Spain}
\affiliation{Institut d'Estudis Espacials de Catalunya (IEEC), E-08034 Barcelona, Spain}
\author[0000-0003-4079-3263]{J.~Frieman}
\affiliation{Fermi National Accelerator Laboratory, P. O. Box 500, Batavia, IL 60510, USA}
\affiliation{Kavli Institute for Cosmological Physics, University of Chicago, Chicago, IL 60637, USA}
\author[0000-0002-9370-8360]{J.~Garc\'ia-Bellido}
\affiliation{Instituto de Fisica Teorica UAM/CSIC, Universidad Autonoma de Madrid, 28049 Madrid, Spain}
\author{E.~Gaztanaga}
\affiliation{Institute of Space Sciences (ICE, CSIC),  Campus UAB, Carrer de Can Magrans, s/n,  08193 Barcelona, Spain}
\affiliation{Institut d'Estudis Espacials de Catalunya (IEEC), 08034 Barcelona, Spain}
\author[0000-0001-6942-2736]{D.~W.~Gerdes}
\affiliation{Department of Astronomy, University of Michigan, Ann Arbor, MI 48109, USA}
\affiliation{Department of Physics, University of Michigan, Ann Arbor, MI 48109, USA}
\author{D.~Gruen}
\affiliation{SLAC National Accelerator Laboratory, Menlo Park, CA 94025, USA}
\affiliation{Kavli Institute for Particle Astrophysics \& Cosmology, P. O. Box 2450, Stanford University, Stanford, CA 94305, USA}
\affiliation{Department of Physics, Stanford University, 382 Via Pueblo Mall, Stanford, CA 94305, USA}
\author{J.~Gschwend}
\affiliation{Laborat\'orio Interinstitucional de e-Astronomia - LIneA, Rua Gal. Jos\'e Cristino 77, Rio de Janeiro, RJ - 20921-400, Brazil}
\affiliation{Observat\'orio Nacional, Rua Gal. Jos\'e Cristino 77, Rio de Janeiro, RJ - 20921-400, Brazil}
\author[0000-0003-0825-0517]{G.~Gutierrez}
\affiliation{Fermi National Accelerator Laboratory, P. O. Box 500, Batavia, IL 60510, USA}
\author{W.~G.~Hartley}
\affiliation{Department of Physics \& Astronomy, University College London, Gower Street, London, WC1E 6BT, UK}
\affiliation{Department of Physics, ETH Zurich, Wolfgang-Pauli-Strasse 16, CH-8093 Zurich, Switzerland}
\author{S.~R.~Hinton}
\affiliation{School of Mathematics and Physics, University of Queensland,  Brisbane, QLD 4072, Australia}
\author{K.~Honscheid}
\affiliation{Center for Cosmology and Astro-Particle Physics, The Ohio State University, Columbus, OH 43210, USA}
\affiliation{Department of Physics, The Ohio State University, Columbus, OH 43210, USA}
\author{E.~Krause}
\affiliation{Department of Astronomy/Steward Observatory, University of Arizona, 933 North Cherry Avenue, Tucson, AZ 85721-0065, USA}
\author[0000-0003-0120-0808]{K.~Kuehn}
\affiliation{Lowell Observatory, 1400 Mars Hill Road, Flagstaff, AZ 86001, USA}
\affiliation{Australian Astronomical Optics, Macquarie University, North Ryde, NSW 2113, Australia}
\author[0000-0003-2511-0946]{N.~Kuropatkin}
\affiliation{Fermi National Accelerator Laboratory, P. O. Box 500, Batavia, IL 60510, USA}
\author{O.~Lahav}
\affiliation{Department of Physics \& Astronomy, University College London, Gower Street, London, WC1E 6BT, UK}
\author{M.~A.~G.~Maia}
\affiliation{Laborat\'orio Interinstitucional de e-Astronomia - LIneA, Rua Gal. Jos\'e Cristino 77, Rio de Janeiro, RJ - 20921-400, Brazil}
\affiliation{Observat\'orio Nacional, Rua Gal. Jos\'e Cristino 77, Rio de Janeiro, RJ - 20921-400, Brazil}
\author[0000-0003-0710-9474]{J.~L.~Marshall}
\affiliation{George P. and Cynthia Woods Mitchell Institute for Fundamental Physics and Astronomy, and Department of Physics and Astronomy, Texas A\&M University, College Station, TX 77843,  USA}
\author[0000-0002-1372-2534]{F.~Menanteau}
\affiliation{Department of Astronomy, University of Illinois at Urbana-Champaign, 1002 West Green Street, Urbana, IL 61801, USA}
\affiliation{National Center for Supercomputing Applications, 1205 West Clark Street, Urbana, IL 61801, USA}
\author[0000-0002-6610-4836]{R.~Miquel}
\affiliation{Instituci\'o Catalana de Recerca i Estudis Avan\c{c}ats, E-08010 Barcelona, Spain}
\affiliation{Institut de F\'{\i}sica d'Altes Energies (IFAE), The Barcelona Institute of Science and Technology, Campus UAB, 08193 Bellaterra (Barcelona) Spain}
\author{A.~Palmese}
\affiliation{Fermi National Accelerator Laboratory, P. O. Box 500, Batavia, IL 60510, USA}
\affiliation{Kavli Institute for Cosmological Physics, University of Chicago, Chicago, IL 60637, USA}
\author{F.~Paz-Chinch\'{o}n}
\affiliation{Department of Astronomy, University of Illinois at Urbana-Champaign, 1002 West Green Street, Urbana, IL 61801, USA}
\affiliation{National Center for Supercomputing Applications, 1205 West Clark Street, Urbana, IL 61801, USA}
\author[0000-0002-2598-0514]{A.~A.~Plazas}
\affiliation{Department of Astrophysical Sciences, Princeton University, Peyton Hall, Princeton, NJ 08544, USA}
\author[0000-0002-9328-879X]{A.~K.~Romer}
\affiliation{Department of Physics and Astronomy, Pevensey Building, University of Sussex, Brighton, BN1 9QH, UK}
\author[0000-0002-9646-8198]{E.~Sanchez}
\affiliation{Centro de Investigaciones Energ\'eticas, Medioambientales y Tecnol\'ogicas (CIEMAT), Madrid, Spain}
\author{B.~Santiago}
\affiliation{Instituto de F\'\i sica, UFRGS, Caixa Postal 15051, Porto Alegre, RS---91501-970, Brazil}
\affiliation{Laborat\'orio Interinstitucional de e-Astronomia - LIneA, Rua Gal. Jos\'e Cristino 77, Rio de Janeiro, RJ - 20921-400, Brazil}
\author{V.~Scarpine}
\affiliation{Fermi National Accelerator Laboratory, P. O. Box 500, Batavia, IL 60510, USA}
\author{S.~Serrano}
\affiliation{Institute of Space Sciences (ICE, CSIC),  Campus UAB, Carrer de Can Magrans, s/n,  08193 Barcelona, Spain}
\affiliation{Institut d'Estudis Espacials de Catalunya (IEEC), 08034 Barcelona, Spain}
\author[0000-0002-3321-1432]{M.~Smith}
\affiliation{School of Physics and Astronomy, University of Southampton,  Southampton, SO17 1BJ, UK}
\author[0000-0001-6082-8529]{M.~Soares-Santos}
\affiliation{Brandeis University, Physics Department, 415 South Street, Waltham, MA 02453, USA}
\author[0000-0002-7047-9358]{E.~Suchyta}
\affiliation{Computer Science and Mathematics Division, Oak Ridge National Laboratory, Oak Ridge, TN 37831, USA}
\author[0000-0003-1704-0781]{G.~Tarle}
\affiliation{Department of Physics, University of Michigan, Ann Arbor, MI 48109, USA}
\author{D.~Thomas}
\affiliation{Institute of Cosmology and Gravitation, University of Portsmouth, Portsmouth, PO1 3FX, UK}
\author{T.~N.~Varga}
\affiliation{Max Planck Institute for Extraterrestrial Physics, Giessenbachstrasse, D-85748 Garching, Germany}
\affiliation{Universit\"ats-Sternwarte, Fakult\"at f\"ur Physik, Ludwig-Maximilians Universit\"at M\"unchen, Scheinerstr. 1, D-81679 M\"unchen, Germany}
\author[0000-0002-7123-8943]{A.~R.~Walker}
\affiliation{Cerro Tololo Inter-American Observatory, National Optical Astronomy Observatory, Casilla 603, La Serena, Chile}

\collaboration{67}{(DES Collaboration)}



\begin{abstract}
The population of Milky Way (MW) satellites contains the faintest known galaxies and thus provides essential insight into galaxy formation and dark matter microphysics. Here we combine a model of the galaxy--halo connection with newly derived observational selection functions based on searches for satellites in photometric surveys over nearly the entire high Galactic latitude sky. In particular, we use cosmological zoom-in simulations of MW-like halos that include realistic Large Magellanic Cloud (LMC) analogs to fit the position-dependent MW satellite luminosity function. We report decisive evidence for the statistical impact of the LMC on the MW satellite population due to an estimated $6\pm 2$ observed LMC-associated satellites, consistent with the number of LMC satellites inferred from Gaia proper-motion measurements, confirming the predictions of cold dark matter models for the existence of satellites within satellite halos. Moreover, we infer that the LMC fell into the MW within the last $2\ \rm{Gyr}$ at high confidence. Based on our detailed full-sky modeling, we find that the faintest observed satellites inhabit halos with peak virial masses below $3.2\times 10^{8}\ M_{\rm{\odot}}$ at $95\%$ confidence, and we place the first robust constraints on the fraction of halos that host galaxies in this regime. We predict that the faintest detectable satellites occupy halos with peak virial masses above $10^{6}\ M_{\rm{\odot}}$, highlighting the potential for powerful galaxy formation and dark matter constraints from future dwarf galaxy searches.
\end{abstract}

\keywords{\href{http://astrothesaurus.org/uat/353}{Dark matter (353)}; \href{http://astrothesaurus.org/uat/1049}{Milky Way dark matter halo (1049)}; \href{http://astrothesaurus.org/uat/574}{Galaxy abundances (574)}}


\section{Introduction}
\label{Introduction}

The sample of confirmed and candidate Milky Way (MW) satellite galaxies has more than doubled in the last $5$ years. Modern imaging surveys have driven these discoveries; in particular, following the successes of the Sloan Digital Sky Survey (SDSS) in the early $2000$s \citep{Willman0410416,Willman0503552,Belokurov0604355,Belokurov0608448,Belokurov08072831,Belokurov09030818,Belokurov10020504,Grillmair0605396,Grillmair08113965,Sakamoto0610858,Zucker0606633,Irwin0701154,Walsh07051378}, the Dark Energy Survey (DES) and the Panoramic Survey Telescope and Rapid Response System Pan-STARRS1 (PS1) have discovered $17$ and three new satellite galaxy candidates, respectively (\citealt*{Bechtol150302584}; \citealt*{Drlica-Wagner150803622}; \citealt{Kim:2015c,Koposov150302079}; \citealt{Laevens150305554,Laevens150707564,Luque:2015}). These systems are identified as arcminute-scale overdensities of individually resolved stars, and many have already been spectroscopically confirmed. Meanwhile, other surveys with the Dark Energy Camera and VST ATLAS have recently discovered several additional satellites (\citealt{martin_2015_hydra_ii,Drlica-Wagner160902148,Torrealba160107178,Torrealba160505338,Torrealba180107279,Koposov180406430,Mau:2019b}).

Nonetheless, the current census of MW satellites is likely highly incomplete, particularly for faint systems in the outer regions of the MW halo. This is evidenced by the detection of three new ultrafaint satellites in the first $\sim 676\ \rm{deg}^2$ of Hyper Suprime-Cam Strategic Survey Program (HSC-SSP) imaging data \citep{Homma160904346,Homma170405977,Homma190607332} and by the discovery of Antlia II, the lowest surface brightness galaxy currently known, using RR Lyrae member stars identified in Gaia DR2 \citep{Torrealba181104082}. In the near future, the Legacy Survey of Space and Time (LSST) conducted from the Vera C.\ Rubin Observatory will be able to detect satellites over the entire southern sky down to a surface brightness of $\mu_V \sim 32\ \rm{mag\ arcsec}^{-2}$ \citep{Ivezic08052366,Tollerud08064381,Hargis14074470,Nadler180905542}.

Interpreting the cosmological and astrophysical implications of these discoveries requires a detailed understanding of the observational selection effects for each survey under consideration. In a companion paper (\citealt*{PaperI}, hereafter \citetalias{PaperI}), we derive observational selection functions for DES and PS1 based on searches for simulated satellites in each dataset. These selection functions encode the probability that satellites in either survey are detectable as a function of their absolute magnitude, heliocentric distance, physical size, and position on the sky. They incorporate realistic photometric error models, selection masks that exclude highly reddened regions near the Galactic disk, and the influence of local stellar density on satellite detectability. Detection sensitivity is linked to sky position because various surveys have imaged different parts of the sky at varying depths, and accurately modeling this effect is crucial in order to disentangle anisotropy in the underlying MW satellite system from selection effects.

In this paper, we combine the observational selection functions derived in \citetalias{PaperI} with a detailed model of the galaxy--halo connection and high-resolution cosmological zoom-in simulations of MW-mass host halos to infer the position-dependent MW satellite luminosity function. Although several empirical models have recently been used to study subsets of the MW satellite population \citep{Jethwa161207834,Kim1812121,NewtonMNRAS,Nadler180905542}, this is the first analysis that is directly based on imaging data over more than $\sim 15,000\ \rm{deg}^2$; indeed, our analysis covers $75\%$ of the high Galactic latitude sky. Moreover, our galaxy--halo connection model allows us to marginalize over astrophysical uncertainties in our fit to the observed DES and PS1 satellite populations. We quantify the impact of the largest MW satellite, the Large Magellanic Cloud (LMC), and its associated satellites on the observed DES and PS1 satellite populations. We find that the satellites accreted with a realistic LMC analog---defined in terms of its mass, heliocentric distance, and infall time---are essential to fit the DES and PS1 luminosity functions simultaneously; this finding constitutes a remarkable confirmation of hierarchical structure formation. We predict that $4.8\pm 1.7$ ($1.1\pm 0.9$) of the known satellites observed by DES (PS1) fell into the MW with the LMC, consistent with the number of LMC-associated satellites inferred from Gaia proper-motion measurements \citep{Kallivayalil_2018,Patel200101746}.

Our analysis constrains the properties of the lowest-mass halos that host observed satellites, which we infer to have peak virial masses below $3.2\times 10^{8}\ M_{\rm{\odot}}$ at $95\%$ confidence. This finding, along with constraints on the faint-end slope of the luminosity function, can be used to inform feedback prescriptions in hydrodynamic simulations \citep{Sawala14066362,Fitts180106187,Simpson170503018,Munshi181012417,Wheeler181202749}. Constraints on the minimum halo mass also hold broad implications for the microphysical properties of dark matter (e.g., \citealt{Drlica-Wagner190201055,Nadler190410000}). Crucially, our model can be extended to explore the degeneracies between baryonic physics and deviations from the cold dark matter (CDM) paradigm.

This paper is organized as follows. We first provide an overview of our framework in Section \ref{overview}. We then describe the simulations (Section \ref{simulations}), galaxy--halo connection model (Section \ref{model}), observational selection functions (Section \ref{osf}), and statistical framework (Section \ref{stats}) used in our analysis. We present our results in Section \ref{results}, focusing on  the observed DES and PS1 satellite populations (\S\ref{obs}), the impact of the LMC system (\S\ref{LMCimpact}), the total MW satellite population (\S\ref{MWpop}), galaxy--halo connection model constraints (\S\ref{constraints}), the properties of halos that host the faintest observed satellites (\S\ref{faint}), and the implications of our findings for dark matter microphysics (\S\ref{microphysics}). We discuss the main theoretical uncertainties in our analysis in Section~\ref{uncertainties}, and we conclude in Section~\ref{discussion}. Appendices provide additional details on our galaxy--halo connection model (Appendix \ref{appendixa}) and statistical framework (Appendix \ref{appendixb}), the robustness of our results to observational systematics (Appendix \ref{appendixd}) and resolution effects (Appendix \ref{appendixd4}), and the observed DES and PS1 satellite populations (Appendix \ref{appendixe}).

Throughout, we use the term ``galaxy--halo connection model'' to refer to a model that describes how the properties of galaxies, including luminosity and size, are related to the properties of halos, such as peak virial mass. Furthermore, ``log'' refers to the base-$10$ logarithm.


\section{Analysis Overview}
\label{overview}

\begin{figure*}[t]
    \centering
    \includegraphics[scale=0.48]{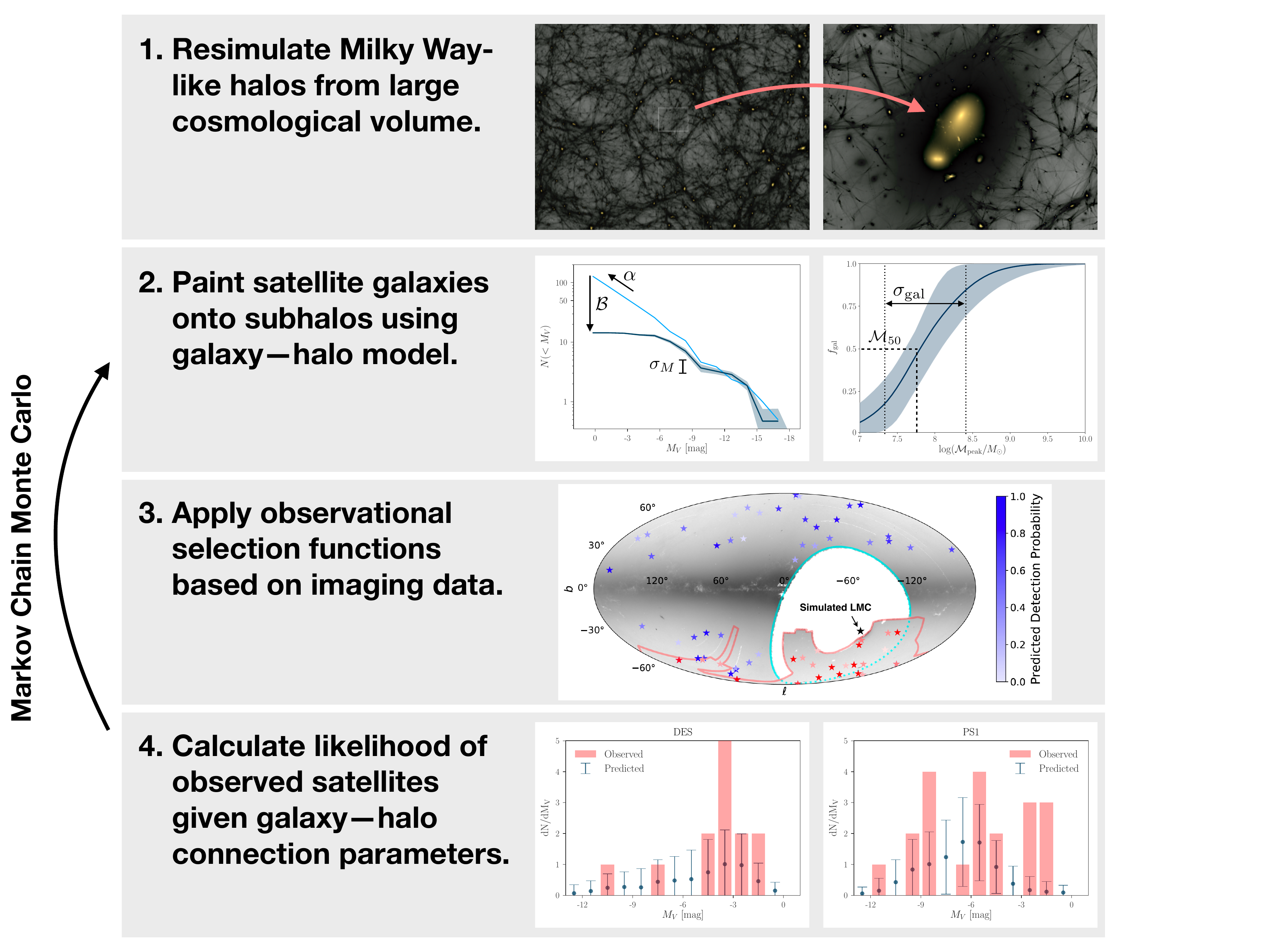}
    \caption{Visualization of our MW satellite modeling framework. In the first step, we perform high-resolution zoom-in simulations of MW-like halos selected from a larger cosmological volume (Section \ref{simulations}); in the second step, we paint galaxies onto subhalos using a parametric model for the galaxy--halo connection (Section \ref{model}); in the third step, we use the observational selection functions derived in \citetalias{PaperI} to compute the probability that these satellites would be observed in DES or PS1 imaging data (Section \ref{osf}); and in the final step, we calculate the likelihood of producing the true DES and PS1 satellite populations given many mock satellite population realizations at fixed galaxy--halo connection model parameters (Section \ref{stats}). We then iterate this process to constrain our galaxy--halo connection model.}
    \label{fig:framework}
\end{figure*}

Using the observed population of MW satellites to constrain our galaxy--halo connection model requires the following ingredients (see Figure \ref{fig:framework} for a visualization of each step).
\begin{enumerate}
\item A model that predicts the \emph{underlying} MW satellite population.
\item An observational selection function that, convolved with the prediction in the previous step, yields a prediction for the \emph{observed} satellite population.
\item A model for the likelihood of producing the true MW satellite population given the prediction from the previous step.
\end{enumerate}

The first step above can either be performed using either a hydrodynamic simulation, in which galaxy formation is modeled at the simulation level, or an empirical prescription for painting galaxies onto halos. We take the latter approach in this paper, which allows for a more flexible modeling framework, as well as the use of simulations that are closer approximations to the MW system. Indeed, our results may help to constrain feedback prescriptions in hydrodynamic simulations. Note that the assumed dark matter model (e.g., cold vs.\ warm dark matter or collisionless vs.\ interacting dark matter) affects the underlying satellite population and often manifests as a cutoff in the abundance of halos---and thus faint galaxies---below a halo mass threshold determined by the microphysical properties of dark matter.

The steps outlined above each rely on tools that have been developed in previous studies and in \citetalias{PaperI}. Here we simply provide a brief description of each step, and we defer additional details to the Appendices.

\section{Simulations}
\label{simulations}

\subsection{General Description}
\label{description}

Our model of the underlying MW satellite population is built on high-resolution dark matter--only zoom-in simulations of MW-mass host halos selected from the suite of~$45$ hosts presented in \cite{Mao150302637}, which have virial masses between $1.2$ and $1.6\times 10^{12}\ M_{\rm{\odot}}$.\footnote{We define virial quantities according to the Bryan--Norman virial overdensity \citep{Bryan_1998}, with $\Delta_{\rm vir}\simeq 99.2$ as appropriate for the cosmological parameters adopted in our zoom-in simulations: $h = 0.7$, $\Omega_{\rm m} = 0.286$, $\Omega_{\rm b} = 0.047$, and $\Omega_{\Lambda} = 0.714$ \citep{Mao150302637}.} The highest-resolution particles in these simulations have a mass of~$3\times 10^{5}\ M_{\rm \odot}\ h^{-1}$, and the softening length in the highest-resolution regions is $170\ \mathrm{pc}\ h^{-1}$.

Halo catalogs and merger trees were generated using the {\sc Rockstar} halo finder and the {\sc consistent-trees} merger code \citep{Behroozi11104372,Behroozi11104370}. Subhalos in these simulations are well resolved down to a present-day maximum circular velocity of $V_{\rm max} \approx 9\ \rm{km\ s}^{-1}$ \citep{Mao150302637}. To be conservative, we only use subhalos with both $V_{\rm max} > 9\ \rm{km\ s}^{-1}$ and peak maximum circular velocity $V_{\rm peak} > 10\ \rm{km\ s}^{-1}$. In Appendix \ref{appendixd4}, we show that these resolution thresholds are sufficient for modeling the satellite populations of interest here.

\subsection{Host Halo and LMC Analog Selection}

The MW might be atypical compared to the average halos of a similar mass (e.g., \citealt{BoylanKolchin09114484,Busha10112203,RodriguezPuebla13064328,Fielder180705180}); in particular, its satellite population is likely affected by the existence of the LMC system and the ``satellites of satellites'' that accreted with the LMC into the virial radius of the MW \citep{Lynden-Bell1976,DOnghia08020001,Lu160502075,Dooley170305321}. In addition, the detailed merger history of the MW---such as the early accretion of an LMC-mass galaxy inferred from Gaia data---might affect its faint satellite population \citep{Bose190904039}.

Thus, we select MW-like host halos that each have an LMC analog with realistic internal and orbital properties; both of these hosts experience an early major merger similar to the Gaia-Enceladus accretion event (see Appendix \ref{appendixa3} for details). We define realistic LMC analogs as subhalos with
\begin{enumerate}
\item present-day maximum circular velocity~$V_{\rm{max}}\geqslant 55\ \rm{km\ s}^{-1}$,
\item present-day heliocentric distance $40\ \mathrm{kpc}<D<60\ \rm{kpc}$, and
\item time of accretion onto the MW less than $2\ \rm{Gyr}$ ago. 
\end{enumerate}
These criteria yield two MW-like host halos with virial masses of $1.57$ and $1.26\times 10^{12}\ M_{\rm{\odot}}$, respectively. Both of these hosts were used in the less restrictive host halo set defined in \cite{Nadler180905542}, and both have a Navarro--Frenk--White (NFW) concentration parameter that is consistent with constraints set using the combination of satellite and globular cluster dynamics measured by Gaia \citep{Callingham180810456,Watkins180411348}. The LMC analogs in these two simulations have present-day virial masses of $1.6$ and $2.5\times 10^{11}\ M_{\rm{\odot}}$ respectively, and both have peak virial masses of $3.0\times 10^{11}\ M_{\rm{\odot}}$. These LMC analogs accreted onto their host halos $1$ and~$1.5\ \rm{Gyr}$ ago, respectively, and their orbital dynamics are consistent with LMC proper-motion measurements (e.g., \citealt{Kallivayalil13010832}).

Our fiducial LMC analogs have masses that are consistent with LMC mass estimates based on stellar stream dynamics, satellite dynamics, and the orbital histories of both Magellanic Clouds \citep{Besla151103346,Penarrubia150703594,Erkal190709484,Erkal181208192}.\footnote{Although detailed exploration of Magellanic Cloud binary systems is beyond the scope of this work, we note that \cite{Shao180307269} and \cite{Cautun180909116} found that LMC analogs with SMC-like companions are typically more massive than isolated LMC analogs.} However, different studies have adopted various definitions of ``LMC mass,’’ and precision in the LMC mass definition (and particularly in the distinction between peak and present-day halo mass) is crucial going forward. We expect that our inference is most sensitive to the peak mass rather than the present-day mass of the LMC because peak mass correlates more directly with the expected abundance of LMC satellites, particularly for recent infall scenarios. Other probes of LMC mass are likely sensitive to these quantities in different ways, and some---including timing arguments (e.g., \citealt{Penarrubia150703594}) and orbit-rewinding to infer LMC satellite abundances (e.g., \citealt{Patel200101746})---might be most sensitive to the ratio of the LMC and MW halo masses.

Although the masses of our host halos are consistent with observational constraints for the MW (e.g., \citealt{Busha10112203,Bland-Hawthorn160207702,Patel170305767}), our simulations span a narrower range of host mass relative to the uncertainty on this quantity inferred from Gaia measurements. For example, \cite{Callingham180810456} found that the MW host virial mass lies between $1.0$ and $1.8\times 10^{12}\ M_{\rm{\odot}}$ at the~$95\%$ confidence level (also see \citealt{Cautun191104557,Li191011257,Li191202086}). Since subhalo abundance is proportional to host halo mass, predicted satellite abundances scale linearly with MW mass, modulo second-order changes in subhalo disruption due to variations in the mass accretion history of the central galaxy \citep{Kelley181112413,Samuel190411508}. Ideally, our analysis would be performed using MW-like host halos---all of which include realistic LMC analogs---that bracket the current range of allowed MW host mass; however, the availability of such MW-like systems is limited by the statistics of our simulations. Thus, we do not marginalize over the full range of allowed MW host masses in this work. We estimate the potential impact of this uncertainty in Appendix \ref{appendixa1}.


\section{Galaxy--Halo Connection Model}
\label{model}

To associate satellite galaxies with subhalos in the simulations described above, we use a modified version of the model developed in \cite{Nadler180905542}. This model parameterizes the relationship between satellite and subhalo properties and the effects of baryonic physics on subhalo populations in flexible ways, which allows us to marginalize over the relevant theoretical uncertainties. Additional model details and tests are presented in Appendix \ref{appendixa}.

\subsection{Satellite Luminosities}

To associate satellite luminosities with subhalos, we follow \cite{Nadler180905542} by employing an abundance-matching procedure that monotonically relates the absolute $V$-band magnitude of satellites, $M_V$, to the peak circular velocity of subhalos, $V_{\rm{peak}}$.\footnote{We perform abundance matching using $V_{\rm{peak}}$ to incorporate the effects of halo assembly bias and to mitigate the impact of subhalo tidal stripping \citep{Reddick12072160,Lehmann151005651}.} This relation is constrained by the GAMA survey \citep{Loveday150501003,Geha170506743} for bright systems ($M_V<-13\ \rm{mag}$) and is extrapolated into the regime of dim satellites by treating the faint-end slope of the satellite luminosity function, $\alpha$, and the lognormal scatter in luminosity at fixed $V_{\rm{peak}}$, $\sigma_M$ as free parameters. We assume that this scatter is lognormal and constant as a function of halo properties in our fiducial analysis; we explore a mass-dependent scatter model in Appendix \ref{appendixa2}.

Our abundance-matching model is a simple, empirical prescription for assigning satellite luminosities that is not designed to capture the complexities of star formation in ultrafaint dwarf galaxies. For example, \cite{Bose180210096} argued that star formation in systems dimmer than $M_V\sim -5\ \rm{mag}$ is effectively shut down by reionization, resulting in two distinct galaxy populations today. While our abundance-matching model is consistent with the current data, which are fit fairly well by a single power-law luminosity function (see \citetalias{PaperI}), it will be valuable to investigate more detailed models of stellar mass growth and to compare against a wider range of observables, including the inferred star formation histories of MW satellites, in future work.

\subsection{Satellite Sizes}

We assign physical sizes to satellites by extrapolating a modified version of the size--virial radius relation from \cite{Kravstov12122980}, which links a galaxy's stellar 3D half-mass radius to its halo's virial radius, into the faint satellite regime. In particular, we set the mean predicted size of each satellite at accretion according to
\begin{equation}
r_{1/2} \equiv \mathcal{A}\left(\frac{R_{\rm{vir}}}{R_0}\right)^n,
\label{eq:size}
\end{equation}
where $\mathcal{A}$ and $n$ are free parameters, $R_{\rm{vir}}$ denotes the subhalo virial radius measured at accretion, and $R_0=10\ \rm{kpc}$ is a normalization constant. Following \cite{Nadler180905542}, we equate the 3D half-mass radii predicted by Equation \ref{eq:size} to azimuthally averaged projected half-light radii; this conversion neglects mass-to-light weighting and projection effects. Nonetheless, this size relation yields reasonable mean sizes when compared to the observed population of classical and SDSS-discovered satellites \citep{Nadler180905542}.

We draw the size of each satellite from a lognormal distribution with a mean given by Equation~\ref{eq:size} and a standard deviation of $\sigma_{\log R}$, which is a free parameter in our model. When fitting the observed satellite populations, we only compare predicted and mock satellites with $r_{1/2}>10\ \rm{pc}$ in order to exclude likely star clusters from the analysis. We explore a more conservative cut of $r_{1/2}>20\ \rm{pc}$ in Appendix \ref{appendixd2}.

The size prescription described above assumes that satellite sizes are fixed after accretion onto the MW. However, post-infall effects such as tidal stripping and tidal heating can shrink or enlarge satellites depending on their orbital histories \citep{Penarrubia08111579,Errani150104968,Fattahi170703898}. In Appendix~\ref{appendixa4}, we show that our key results are not sensitive to these effects using a simple model for satellite size evolution due to tidal stripping.

\subsection{Subhalo Disruption Due to Baryonic Effects}

To incorporate the effects of baryonic physics---and particularly the tidal influence of the Galactic disk---on our simulated subhalo populations, we apply a random forest algorithm trained on hydrodynamic simulations to predict the probability that each subhalo will be disrupted in a hydrodynamic resimulation based on its orbital and internal properties \citep{Garrison-Kimmel170103792,Nadler171204467}. We model the strength of this disruption effect using the free parameter $\mathcal{B}$, which is defined such that $\mathcal{B}=1$ corresponds to fiducial hydrodynamic predictions \citep{Nadler171204467} and larger (smaller) values of $\mathcal{B}$ correspond to more effective (less effective) subhalo disruption. For each subhalo, we set
\begin{equation}
    p_{\mathrm{disrupt}} \equiv (p_{\mathrm{disrupt},0})^{1/\mathcal{B}}, \label{eq:pdisrupt}
\end{equation}
where $p_{\mathrm{disrupt},0}$ is the fiducial disruption probability returned by the machine-learning algorithm in \cite{Nadler171204467}.

\subsection{Galaxy Formation Efficiency}

The stochastic, nonlinear nature of galaxy formation in low-mass halos likely leads to a smoothly varying fraction of occupied halos, rather than a sharp cutoff in the efficiency of galaxy formation \citep{Sawala14066362,Fitts180106187,Munshi181012417,Wheeler181202749}. Thus, in our fiducial model, we parameterize the fraction of halos that host galaxies of any mass, referred to as the \emph{galaxy occupation fraction}, following \cite{Graus180803654},
\begin{equation}
    f_{\rm{gal}}(\mathcal{M}_{\rm{peak}}) \equiv \frac{1}{2}\Big[1+\mathrm{erf}\Big(\frac{\mathcal{M}_{\rm{peak}}-\mathcal{M}_{50}}{\sqrt{2}\sigma_{\rm{gal}}}\Big)\Big],\label{eq:fgal}
\end{equation}
where $\mathcal{M}_{\rm{peak}}$ is the largest virial mass a subhalo ever attains, which typically occurs before infall into the MW; $\mathcal{M}_{50}$ is the peak halo mass at which $50\%$ of halos host galaxies of any mass; and $\sigma_{\rm{gal}}$
is the width of the galaxy occupation fraction. In our fiducial model, $\mathcal{M}_{50}$ and $\sigma_{\rm{gal}}$ are free parameters. Note that in the limit $\sigma_{\rm{gal}}\rightarrow 0$, this reduces to a model in which all halos with $\mathcal{M}_{\rm{peak}}>\mathcal{M}_{50}$ host a galaxy.

Although our analysis does not constrain $\sigma_{\rm{gal}}$, Equation \ref{eq:fgal} is a simple, physically motivated form of the occupation fraction that will be interesting to explore in future work. Note that we parameterize the occupation fraction in terms of peak halo mass (rather than, e.g., $V_{\rm{peak}}$) because $\mathcal{M}_{\rm{peak}}$ is more easily interpretable and connects directly to constraints on alternative dark matter models (e.g., \citealt{Nadler190410000}).

\subsection{Orphan Satellites}

Because we model faint galaxies that can potentially inhabit subhalos near the resolution limit of our simulations, it is important to account for artificially disrupted subhalos that might host ``orphan'' satellites \citep[][and see \citealt{Bose190904039} for a recent example of the importance of orphans in satellite modeling]{Guo10060106}. To model orphans, we follow the prescription in \cite{Nadler180905542}, which identifies disrupted subhalos in each simulation, interpolates their orbits to $z=0$ using a softened gravitational force law and a dynamical friction model, and accounts for tidal stripping with a mass-loss model calibrated on high-resolution test simulations. We parameterize the effective abundance of orphans by setting their disruption probabilities equal to
\begin{equation}
    p_{\rm{disrupt}} \equiv (1 - a_{\rm acc})^{\mathcal{O}},
\end{equation}
where $a_{\rm{acc}}$ is the final scale factor at which each subhalo enters the virial radius of the MW, and $\mathcal{O}$ is a parameter that captures deviations from disruption probabilities in hydrodynamic simulations, which are well fit by $\mathcal{O}=1$ \citep{Nadler180905542}. Thus, larger (smaller) values of $\mathcal{O}$ correspond to a greater (smaller) contribution from orphan satellites.

Following \cite{Nadler180905542}, we include orphan satellites by fixing $\mathcal{O}=1$ in our fiducial model. Thus, we effectively assume that subhalo disruption in dark matter--only simulations is a numerical artifact \citep{VandenBosch180105427,VandenBosch171105276} but that subhalo disruption in hydrodynamic simulations is a physical effect. We show that our results are largely insensitive to the value of $\mathcal{O}$ in Appendix~\ref{appendixa6}.

\section{Observational Selection Functions}
\label{osf}

We employ the DES and PS1 survey selection functions derived in \citetalias{PaperI}, which have been publicly distributed as machine-learning classifiers that predict satellite detection probability given absolute magnitude, $M_V$; heliocentric distance, $D$; azimuthally averaged projected half-light radius,~$r_{1/2}$; and sky position.\footnote{The DES Y3A2 and PS1 DR1 selection functions are publicly available at \href{https://github.com/des-science/mw-sats}{https://github.com/des-science/mw-sats}.} The predicted detection probabilities are derived from searches for simulated satellites in catalog-level DES and PS1 data, and they employ geometric cuts that restrict observable satellites to lie within the respective survey footprint and that mask regions where satellite detection is challenging due to interstellar extinction, bright nearby stars, and bright extragalactic objects.

We self-consistently apply these position-dependent detection criteria to our predicted satellite populations by matching the on-sky position of our LMC analogs to the true on-sky position of the LMC. In particular, we choose random observer locations $8\ \rm{kpc}$ from the halo center, and we perform appropriate rotations to our subhalo populations for each observer location to match the true LMC position. We apply the DES selection function for satellites within the overlap region of the two surveys, and we only count satellites within a fiducial heliocentric distance of $300\ \rm{kpc}$.

\section{Statistical Framework}
\label{stats}

To fit our galaxy--halo model to the DES and PS1 luminosity functions derived in \citetalias{PaperI}, we generate predicted satellite populations given a set of galaxy--halo connection model parameters, $\boldsymbol{\theta}$, by performing mock DES-plus-PS1 surveys using the selection functions described above. For each host halo and each realization of our satellite model, we bin mock-observed satellites according to their absolute magnitude. We further split satellites in each absolute magnitude bin into high ($\mu_V < 28\ \rm{mag\ arcsec}^{-2}$) and low ($\mu_V \geqslant 28\ \rm{mag\ arcsec}^{-2}$) surface brightness samples to incorporate satellite size information in our fit.\footnote{In particular, we calculate the effective surface brightness averaged within the half-light radius as $\mu_V = M_V + 36.57 + 2.5\log(2\pi r_{1/2}^{2})$, where $r_{1/2}$ is measured in units of $\rm{kpc}$.} We list the DES and PS1 satellites used in this analysis in Table \ref{fig:des_table}.

Next, we calculate the number of predicted satellites in each bin $i$ via
\begin{equation}
    n_i = \sum_{s_i} p_{\mathrm{detect},s_i}\times(1-p_{\mathrm{disrupt},s_i})\times f_{\mathrm{gal},s_i},\label{eq:nsat}
\end{equation}
where $s_i$ indexes the satellites in bin $i$, $p_{\rm{detect}}$ is the detection probability returned by the appropriate observational selection function, $p_{\rm{disrupt}}$ is the disruption probability due to baryonic effects (Equation \ref{eq:pdisrupt}), and $f_{\rm{gal}}$ is the galaxy occupation fraction (Equation \ref{eq:fgal}). For objects that lie in the overlap region of the DES and PS1 footprints, we calculate $p_{\rm{detect}}$ using the DES selection function.

We note that detection probability mainly depends on surface brightness and present-day heliocentric distance (see \citetalias{PaperI}), disruption probability mainly depends on orbital properties \citep{Nadler171204467}, and galaxy occupation depends on~$\mathcal{M}_{\rm{peak}}$ according to Equation \ref{eq:fgal}. Thus, our model for satellite detectability is coupled to our galaxy occupation fraction model, since surface brightness is directly linked to $\mathcal{M}_{\rm{peak}}$ due to our abundance-matching assumption. Nonetheless, our results are largely unaffected if we exclude the galaxy occupation fraction from our model, confirming that $f_{\rm{gal}}$ can be interpreted as the probability that a halo hosts a satellite brighter than $M_V=0\ \rm{mag}$, corresponding to the faintest satellite in our observational sample.

\begin{figure*}[t]
    \includegraphics[scale=0.35]{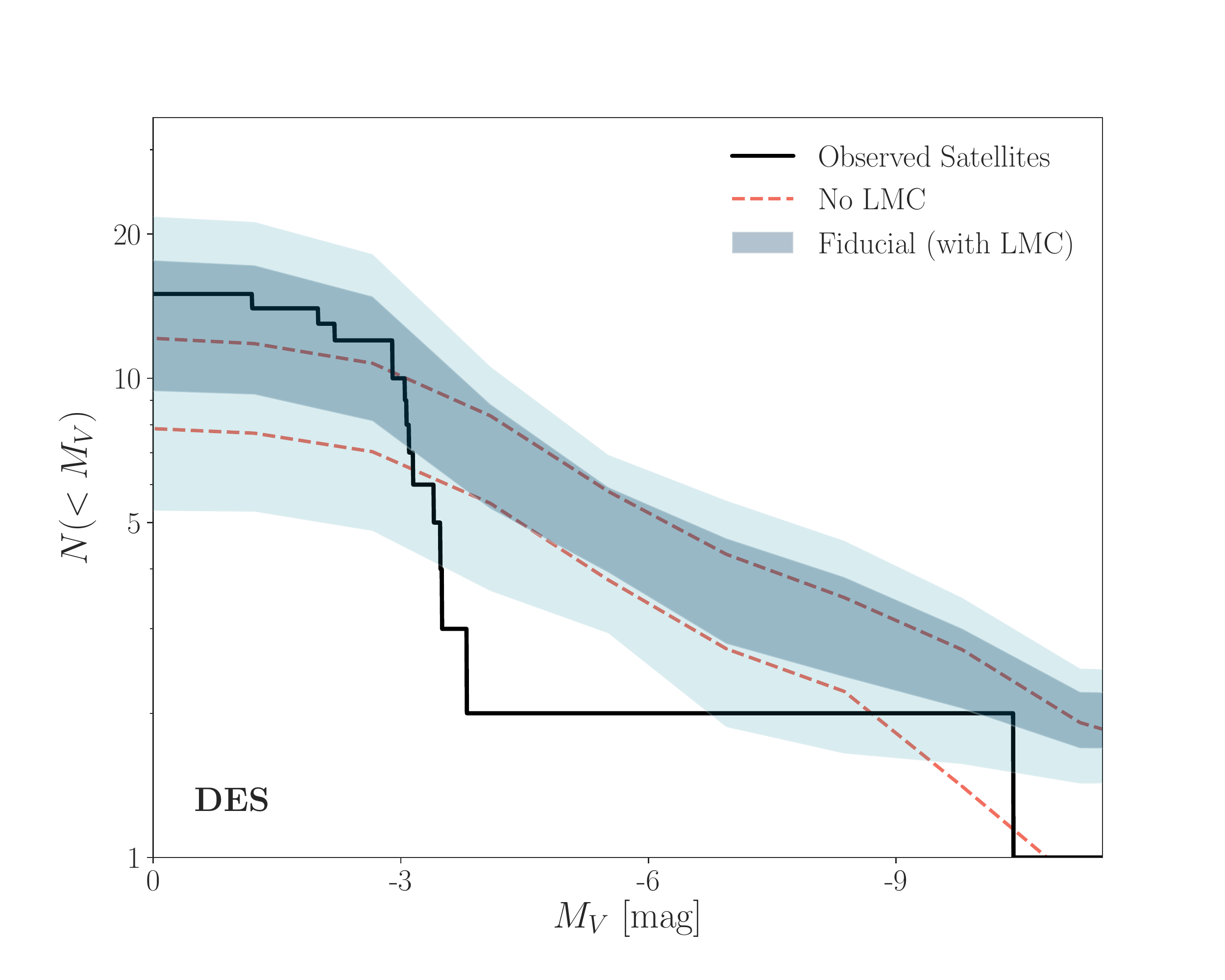}
    \includegraphics[scale=0.35]{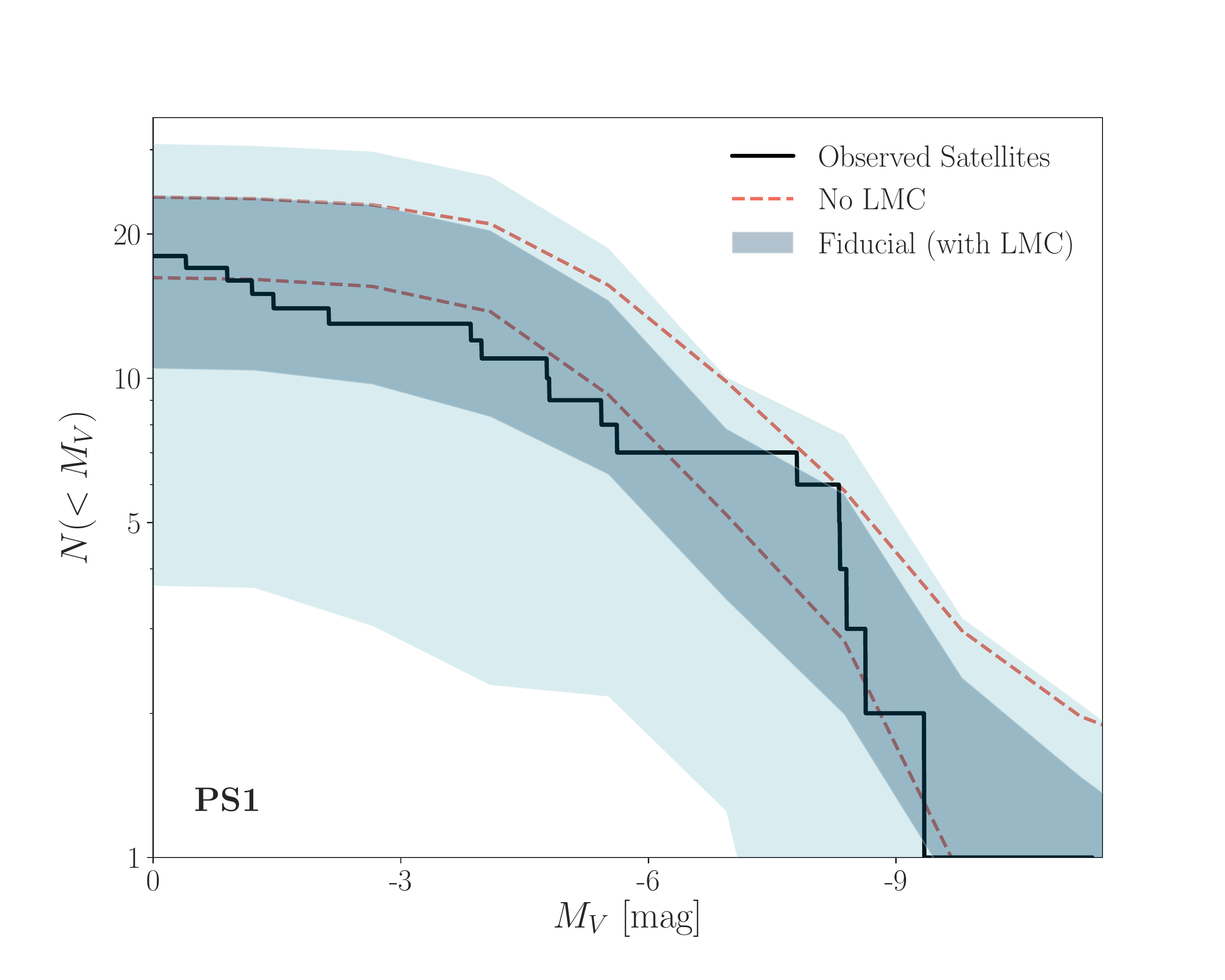}
    \caption{Predicted DES and PS1 satellite luminosity functions resulting from a joint fit to these satellite populations. Dark (light) blue bands correspond to $68\%$ ($95\%$) confidence intervals from our fiducial eight-parameter galaxy--halo connection model, dashed red lines show the $68\%$ confidence interval for a model using host halos without LMC analogs (``No LMC''), and black lines show the observed luminosity functions within each survey footprint. Our fiducial model, which includes realistic LMC analogs, is decisively favored over the No LMC scenario, with a Bayes factor of $\sim 10^4$.}
    \label{fig:LFs}
\end{figure*}

We assume that satellites populate each bin in absolute magnitude--surface brightness parameter space according to an independent Poisson point process with a rate parameter $\lambda$ that depends on absolute magnitude, surface brightness, and our galaxy--halo connection model parameters. Because our model yields noisy estimates of $\lambda$, we marginalize over its range of possible values in each bin, following \cite{Nadler180905542}. The likelihood of observing the set of DES and PS1 satellites, $\mathbf{s}_{\mathrm{DES}}$ and $\mathbf{s}_{\mathrm{PS1}}$ (specified by their absolute magnitudes and surface brightnesses), given a set of model parameters $\boldsymbol{\theta}$ is then
\begin{equation}
    P(\mathbf{s}_{\mathrm{DES}},\mathbf{s}_{\mathrm{PS1}}|\boldsymbol{\theta})=\prod_{\mathrm{  bins\ } i} P(n_{\mathrm{DES},i}|\mathbf{\hat{n}}_{\mathrm{DES},i})\times P(n_{\mathrm{PS1},i}|\mathbf{\hat{n}}_{\mathrm{PS1},i}),\label{eq:likelihood}
\end{equation}
where $n_{\mathrm{DES},i}$ ($n_{\mathrm{PS1},i}$) is the observed number of DES (PS1) satellites in bin $i$, and $\mathbf{\hat{n}}_{\mathrm{DES},i}$ ($\mathbf{\hat{n}}_{\mathrm{PS1},i}$) is a vector of the number of mock DES (PS1) satellites in bin $i$ from several realizations of our model at fixed $\boldsymbol{\theta}$. These realizations include draws over host halos, observer locations, and our galaxy--halo connection model, which is stochastic at fixed $\boldsymbol{\theta}$. Note that steps 1--3 in Figure \ref{fig:framework} generate mock satellite populations $\mathbf{\hat{n}}_{\mathrm{DES}}$ and $\mathbf{\hat{n}}_{\mathrm{PS1}}$, and step 4 compares these to the observed populations $n_{\mathrm{DES}}$ and $n_{\mathrm{PS1}}$.
The explicit forms of~$P(n_{\mathrm{DES},i}|\mathbf{\hat{n}}_{\mathrm{DES},i})$ and $P(n_{\mathrm{PS1},i}|\mathbf{\hat{n}}_{\mathrm{PS1},i})$ are given in Equation~\ref{eq:like}.

Finally, we use Bayes's theorem to compute the posterior distribution over galaxy--halo connection model parameters,
\begin{equation}
P(\boldsymbol{\theta}|\mathbf{s}_{\mathrm{DES}},\mathbf{s}_{\mathrm{PS1}}) = \frac{P(\mathbf{s}_{\mathrm{DES}},\mathbf{s}_{\mathrm{PS1}}|\boldsymbol{\theta})P(\boldsymbol{\theta})}{P(\mathbf{s}_{\mathrm{DES}},\mathbf{s}_{\mathrm{PS1}})},
\end{equation}
where $P(\boldsymbol{\theta})$ is our prior on the galaxy--halo connection model parameters (given in Appendix \ref{appendixb2}), $P(\mathbf{s}_{\mathrm{DES}},\mathbf{s}_{\mathrm{PS1}})$ is the Bayesian evidence, and $P(\mathbf{s}_{\mathrm{DES}},\mathbf{s}_{\mathrm{PS1}}|\boldsymbol{\theta})$
is given by Equation \ref{eq:likelihood}. To sample from this posterior, we run $10^{5}$ iterations of the Markov Chain Monte Carlo (MCMC) sampler \texttt{emcee} \citep{emcee} to sample the eight free parameters $\boldsymbol{\theta} = (\alpha,\sigma_M,\mathcal{M}_{50},\mathcal{B},\sigma_{\rm{gal}},\mathcal{A},\sigma_{\log R},n)$ using $32$ walkers. We discard a burn-in period of $20$ autocorrelation lengths, corresponding to~$\sim 10^4$ steps, which leaves more than~$100$ independent samples.


\section{Results} 
\label{results}

We now present our results, focusing on the observed DES and PS1 satellite populations (\S\ref{obs}), the impact of the LMC system~(\S\ref{LMCimpact}), the total MW satellite population~(\S\ref{MWpop}), the galaxy--halo connection model constraints (\S\ref{constraints}), the properties of the halos that host faint satellites~(\S\ref{faint}), and the implications for dark matter microphysics~(\S\ref{microphysics}).

\begin{figure*}[t]
    \includegraphics[scale=0.35]{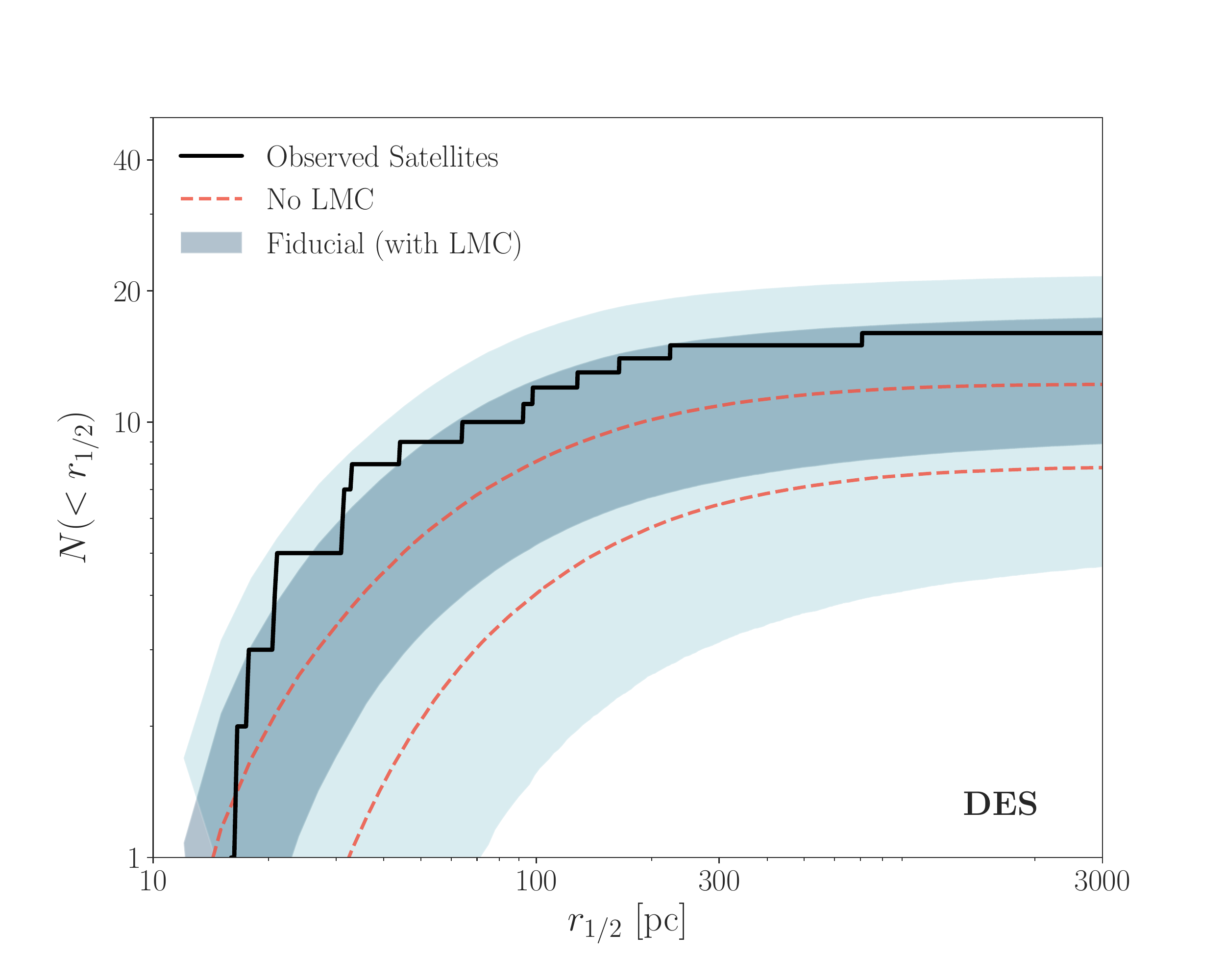}
    \includegraphics[scale=0.35]{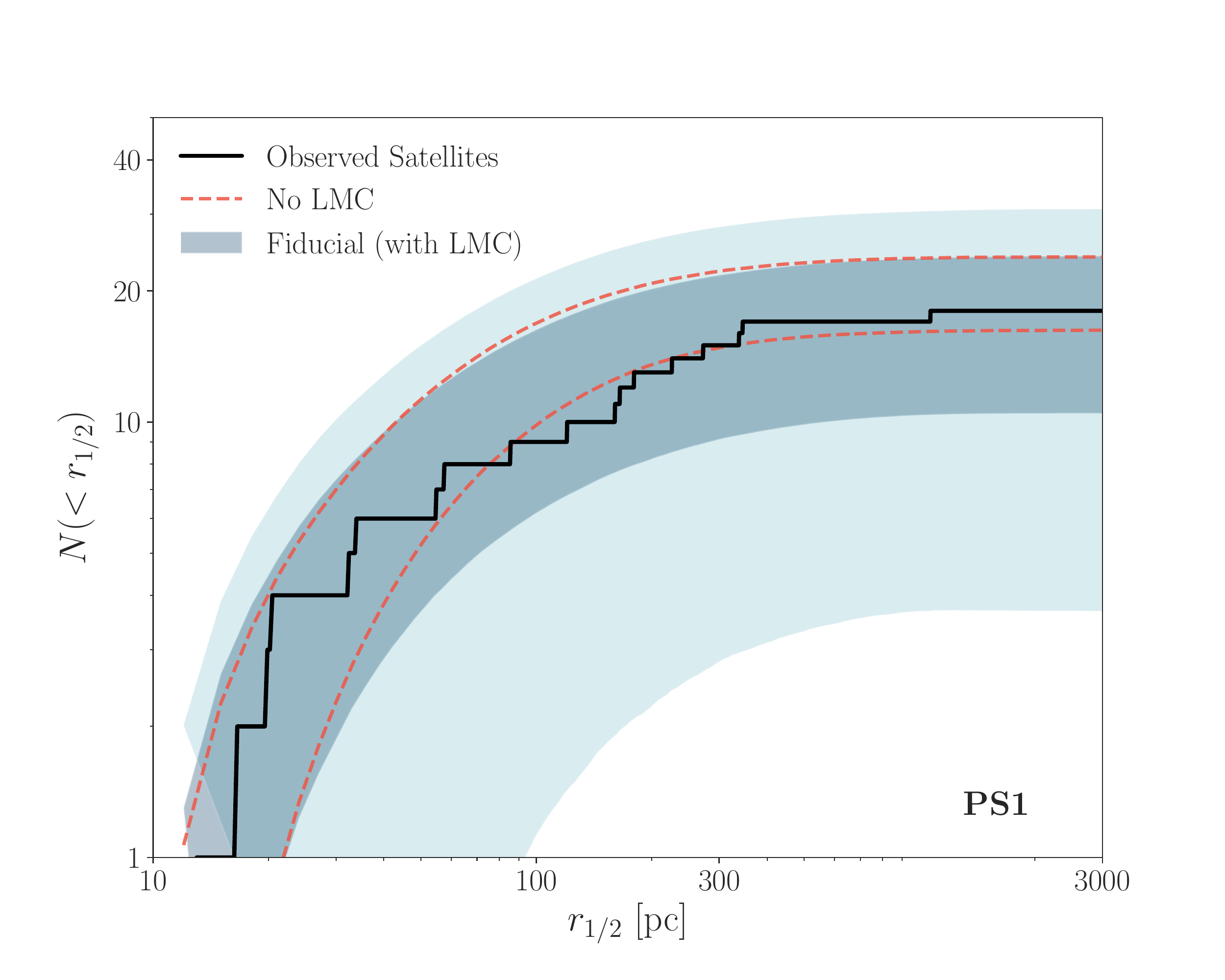}
    \caption{Size distributions derived by fitting our galaxy--halo connection model to the DES and PS1 satellite populations. Dark (light) blue bands correspond to $68\%$ ($95\%$) confidence intervals from our fiducial eight-parameter model, dashed red lines show the $68\%$ confidence interval for a model using host halos without LMC analogs (``No LMC''), and black lines show the observed size distributions.}
    \label{fig:size_dists}
\end{figure*}

\subsection{Observed Satellite Populations}
\label{obs}

Figure \ref{fig:LFs} shows the $68\%$ and $95\%$ confidence intervals for the observed DES and PS1 satellite luminosity functions given by draws from the posterior of our fiducial model, which is consistent with both datasets. We note that the DES and PS1 likelihoods individually yield consistent results.

Figure \ref{fig:size_dists} shows the corresponding satellite size distributions drawn from our fiducial posterior. Our model is consistent with the sizes of both the observed DES and PS1 satellites. It very slightly overpredicts the sizes of observed DES systems; however, we reiterate that our size model does not allow for size reduction due to tidal stripping or size enlargement due to tidal heating, which affect satellites with close pericentric passages to the Galactic disk (e.g., \citealt{Amorisco190105460}). Our findings in Appendix \ref{appendixa4} suggest that the post-infall size evolution of satellites in subhalos with $V_{\rm{peak}}> 10\ \rm{km\ s}^{-1}$ and~$V_{\rm{max}}> 9\ \rm{km\ s}^{-1}$ does not significantly affect our inference.

Our fiducial model is consistent with the outer radial distributions of both DES and PS1 satellites, but it slightly underpredicts the number of satellites near the center of the MW ($D\lesssim 100\ \rm{kpc}$), particularly in PS1. We explore this minor discrepancy in Appendix \ref{appendixa3}, where we show that our galaxy--halo connection model constraints and inferred total MW satellite population are largely unaffected if the radial distribution is forced to match the data.

\subsection{The Impact of the LMC}
\label{LMCimpact}

To assess the impact of the LMC and its satellites on the MW satellite population, we test the following models in addition to our fiducial model, which includes a realistic, recently accreted LMC system by construction.
\begin{enumerate}
\item No LMC: a model with four host halos that have the same mean concentration as our fiducial hosts but no LMC analog.
\item Misplaced LMC: a model with our fiducial host halos in which subhalo positions are reflected, effectively placing the DES footprint in the northern hemisphere.
\item Early LMC Infall: a model with two host halos that have the same mean concentration as our fiducial hosts with LMC analogs that pass our LMC $V_{\rm{max}}$ and heliocentric distance cuts but fall into the MW $2$ and $6\ \rm{Gyr}$ ago, respectively.
\end{enumerate}
For each alternative LMC scenario listed above, we refit the observed DES and PS1 satellite populations, sampling over the same eight parameters used in our fiducial analysis.

Our fiducial model is favored over the No LMC, Misplaced LMC, and Early LMC Infall scenarios with Bayes factors of~$\sim 10^4$, $10^4$, and $10^3$, respectively. In addition, both host halos in the Early LMC Infall case are individually disfavored with Bayes factors of~$\sim 10^ 3$. Thus, we find decisive statistical evidence for the impact of the LMC on the MW satellite population, particularly within and near the DES footprint. Moreover, we infer that the LMC system fell into the MW within the last $2\ \rm{Gyr}$ at high confidence. We also note that, in our fiducial host with more massive MW and LMC halos, the LMC reaches pericenter near the second-to-last simulation snapshot (i.e., $\sim 150\ \rm{Myr}$ ago). Performing our analysis using the final snapshot for this host noticeably degrades the fit due to the dispersal and disruption of LMC satellites during the LMC's pericentric passage. Thus, we use the second-to-last snapshot for this host in our fiducial analysis, and we remark that satellite abundances can potentially constrain the number of allowed pericentric passages for the LMC.

The alternative LMC scenarios defined above are strongly disfavored relative to our fiducial model because they cannot produce a sufficient number of dim satellites in the DES footprint without overpredicting the number of observed PS1 satellites. This is a direct consequence of the spatial overdensity of subhalos near the LMC analogs in our fiducial simulations; in particular, the projected density of resolved subhalos within $50^{\circ}$ of the LMC on the sky is enhanced by~$\sim 50\%$ relative to the density on a random patch of sky.

\begin{figure*}[t]
    \includegraphics[scale=0.35]{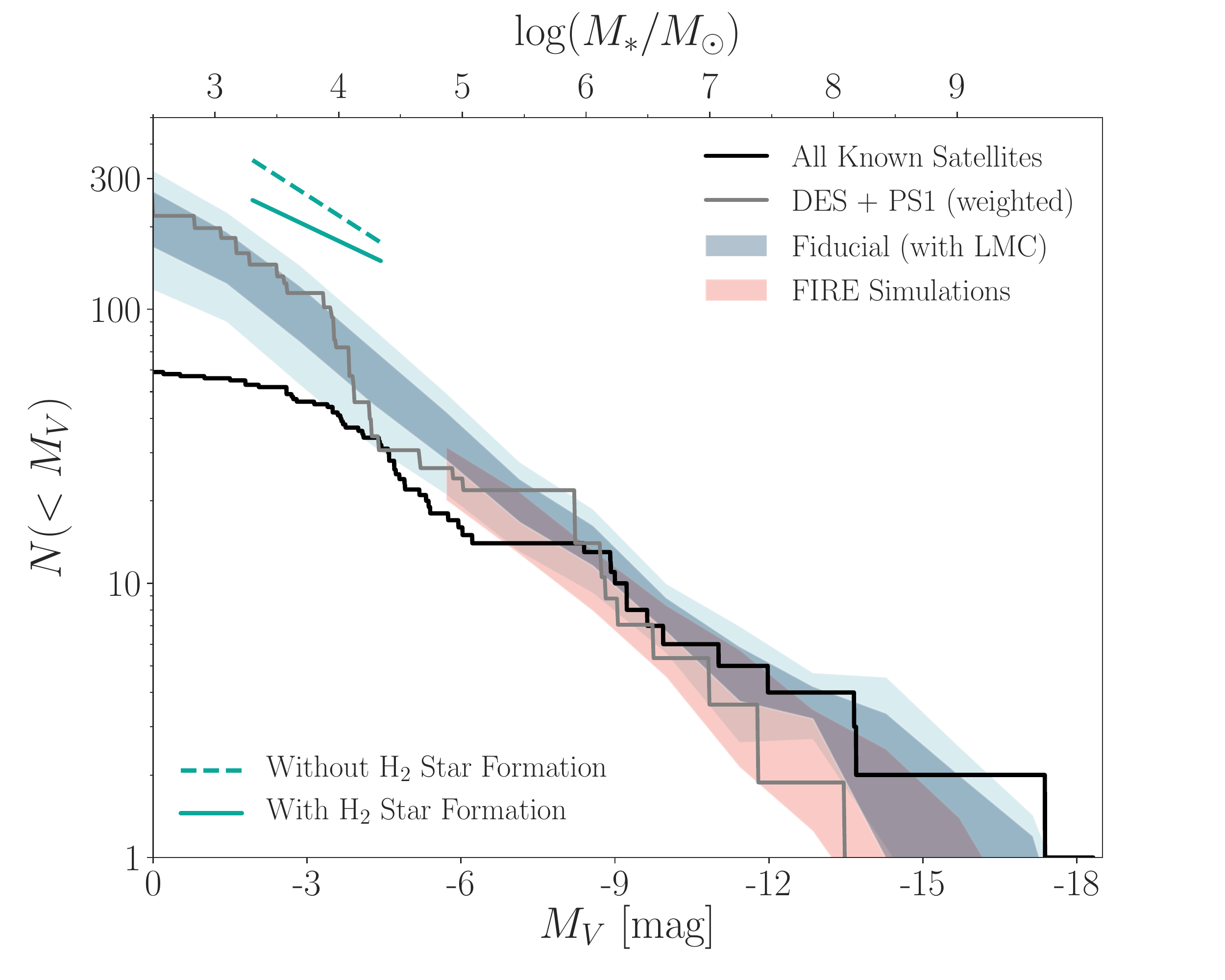}
    \includegraphics[scale=0.35]{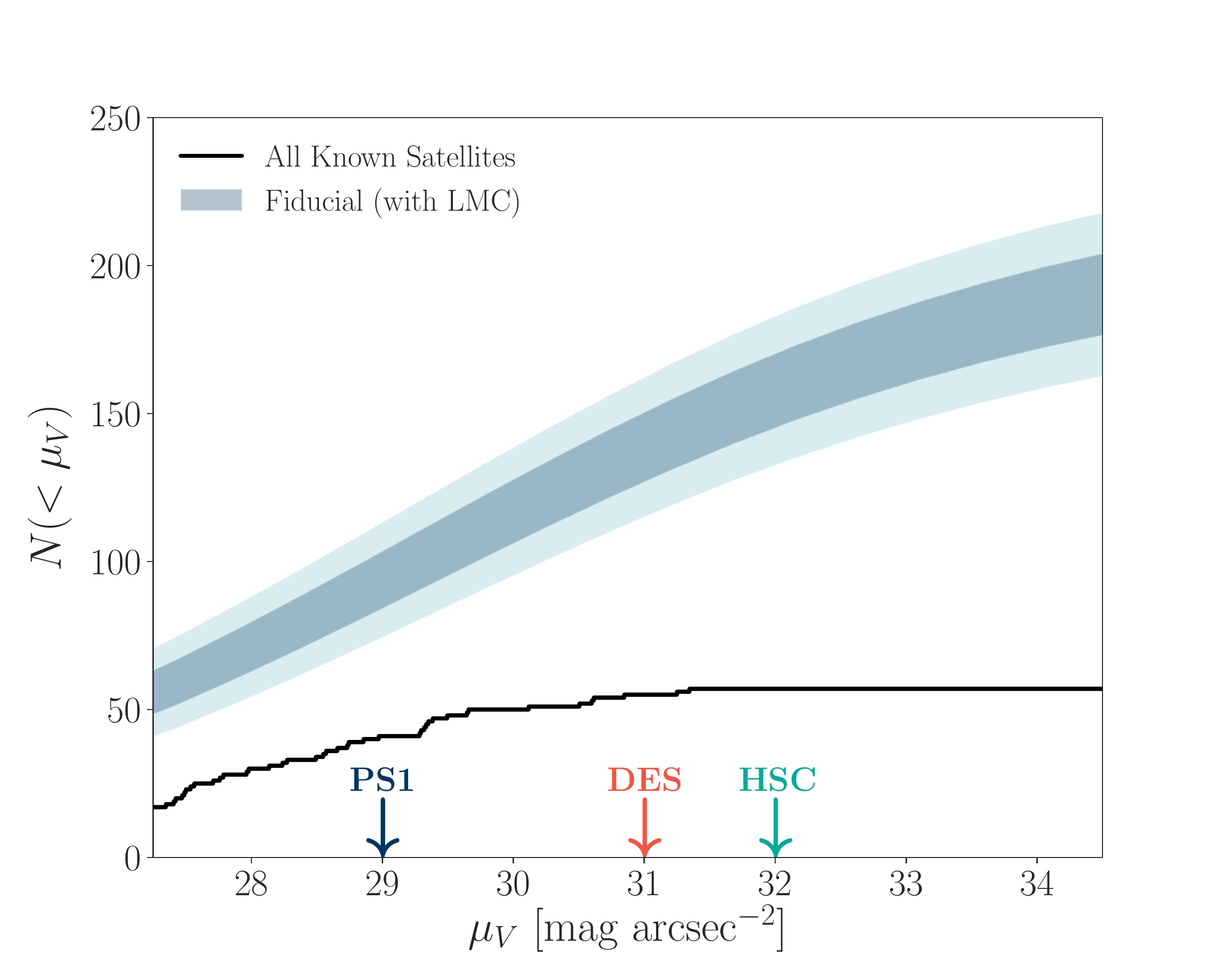}
    \caption{Left panel: total MW satellite luminosity function inferred from our joint fit to the DES and PS1 satellite populations (blue) compared to the current census of confirmed and candidate MW satellites (black) and the empirical estimate derived in \citetalias{PaperI} (gray), which assumes an isotropic satellite distribution and a cored NFW radial satellite distribution. The $68\%$ confidence intervals from hydrodynamic simulations of the Local Group using the FIRE feedback prescription are shown in red \citep{Garrison-Kimmel180604143}. Luminosity function slopes predicted from hydrodynamic simulations with (solid green line) and without (dashed green line) $\rm{H}_2$-based star formation are shown for comparison \citep{Munshi181012417}; these predictions do not account for subhalo disruption due to the Galactic disk. Note that the \citetalias{PaperI} prediction (gray) differs from the ``All Known Satellites'' curve (black) at the bright end because it does not include the LMC, SMC, or Sagittarius. Right panel: The surface brightness distribution of MW satellites with $M_V<0\ \rm{mag}$ and $r_{1/2}>10\ \rm{pc}$ as a function of the limiting observable surface brightness of an all-sky survey. Arrows indicate approximate detection limits for current surveys. Note that LSST Y1 is expected to have similar detection sensitivity to HSC \citep{Ivezic08052366,Tollerud08064381,Hargis14074470,Nadler180905542}.}
    \label{fig:tot_LFs}
\end{figure*}

To quantify the number of satellites in our fiducial model that are associated with the LMC, we explore the following definitions of LMC-associated subhalos.
\begin{enumerate}
\item Fiducial definition. A subhalo is associated with the LMC if it is within the virial radius of the LMC halo at the time of LMC infall into the MW.
\item Gravitationally influenced definition. A subhalo is associated with the LMC if it has ever passed within the virial radius of the LMC halo.
\end{enumerate}
Here LMC infall is defined as the snapshot at which the center of the LMC halo crosses within the MW virial radius. Note that nearly all systems that satisfy our strict definition are bound to the LMC at the time of LMC infall.

Under the fiducial (gravitationally influenced) definitions above, we predict that~$52\pm 8$ ($181\pm 25$) total LMC-associated subhalos (above our cuts of $V_{\rm peak} > 10\ \rm{km\ s}^{-1}$ and $V_{\rm max} > 9\ \rm{km\ s}^{-1}$) exist within the virial radius of the MW today, where the $95\%$ confidence interval is estimated by drawing from our fiducial posterior. We predict that $48\pm 8$ ($164\pm 25$) of these subhalos form galaxies with~$M_V<0\ \rm{mag}$ and $r_{1/2}>10\ \rm{pc}$ (in agreement with an earlier estimate by \citealt{Jethwa161207834}), and that $41\pm 7$ ($118\pm 21$) of these satellites survive tidal disruption due to the Galactic disk. Of these surviving LMC-associated satellites, we predict that~$4.8\pm 1.7$ ($11\pm 3.6$) are currently observed by DES and that $1.1\pm 0.9$ ($6.1\pm 2.1$) are currently observed by PS1.

Our statistical probe of LMC satellite association is remarkably consistent with the number of observed LMC satellites inferred from Gaia proper-motion measurements, which indicate that four galaxies in or near the DES footprint---excluding the Small Magellanic Cloud (SMC)---are associated with the LMC, and that two satellites in or near the PS1 footprint are potentially associated with the LMC (\citealt{Kallivayalil_2018, Patel200101746}).\footnote{A recent analysis based on Gaia proper-motion measurements and hydrodynamic simulations suggests that two bright satellites in or near DES, Fornax and Carina, may also be LMC-associated (\citealt{Pardy190401028}; however, see \citealt{Patel200101746})}. In addition, the orbital dynamics of our predicted LMC satellites are consistent with Gaia proper-motion measurements for these likely LMC-associated systems. These predictions are also consistent with other empirical models \citep{Deason150404372,Jethwa160304420,Dooley170305321,Sales160503574,Kallivayalil_2018,Erkal190709484,Zhang190404296} and with hydrodynamic simulations of isolated LMC analogs \citep{Jahn190702979}.

In Appendix \ref{appendixd4}, we show that the properties of our LMC-like systems are not significantly affected by the realizations of small-scale power in our fiducial simulations. However, we caution that the number of predicted LMC satellites observed by DES and PS1 depends on the particular properties of our two LMC analogs. Thus, exploring the robustness of these results using a suite of zoom-in simulations selected to contain realistic LMC systems with a range of internal and orbital properties is an important avenue for future work.

\begin{figure*}[ht]
\centering
    \includegraphics[scale=0.61]{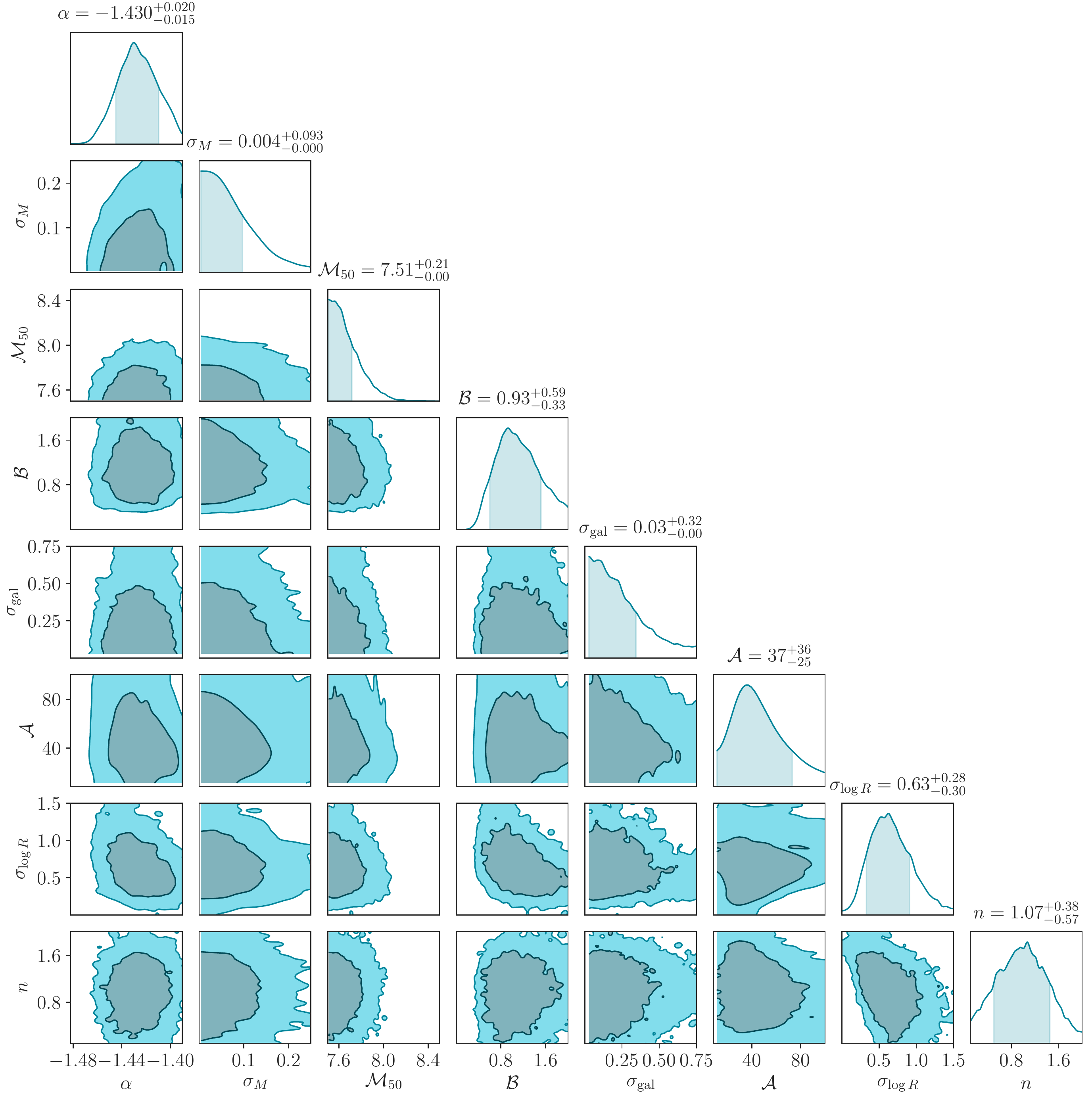}
    \caption{Posterior distribution from our fit to the DES and PS1 satellite populations. Dark (light) shaded contours represent $68\%$ ($95\%$) confidence intervals. Shaded areas in the marginal distributions and parameter summaries correspond to $68\%$ confidence intervals. Note that $\sigma_M$, $\sigma_{\rm{gal}}$, and $\sigma_{\log R}$ are reported in $\rm{dex}$, $\mathcal{M}_{50}$ is reported as $\log(\mathcal{M}_{50}/M_{\rm{\odot}})$, and $\mathcal{A}$ is reported in $\rm{pc}$.}
    \label{fig:posterior}
\end{figure*}

\subsection{The Total MW Satellite Population}
\label{MWpop}

Figure \ref{fig:tot_LFs} shows the total MW satellite luminosity function and surface brightness distribution resulting from our fit to the DES and PS1 satellite populations. We infer that a total of $220\pm 50$ satellites with $M_V<0\ \rm{mag}$ and $r_{1/2}>10\ \rm{pc}$ exist within the virial radius of the MW, where uncertainties correspond to $68\%$ confidence intervals calculated by sampling from our fiducial posterior. Thus, we predict that $\sim 150$ satellites remain undiscovered in a standard CDM scenario, roughly one-fourth of which are associated with the LMC. This is larger than the fraction of satellites that have ever fallen into the MW that are associated with the LMC because our fiducial LMC analogs accreted recently, making their satellites less likely to be disrupted. Our prediction for the total number of MW satellites is consistent with several recent studies \citep{Jethwa161207834,Kim1812121,NewtonMNRAS,Nadler180905542}, and it is lower than the empirical estimate in \citetalias{PaperI}, which was recognized to be inflated due to the assumption of an isotropic satellite distribution. This prediction will be tested by upcoming deep imaging surveys; indeed, HSC-SSP has already started to probe this population of distant, low surface brightness MW satellites by discovering three new ultrafaint satellite candidates in $\sim 676\ \rm{deg}^2$ of imaging data \citep{Homma160904346,Homma170405977,Homma190607332}.

To estimate whether our predictions are consistent with HSC-SSP observations, we draw realizations of the MW satellite population from our fiducial posterior and calculate the number of systems within the DES or PS1 footprints that would not be observed by the appropriate survey. We then estimate the number of these systems currently observed by an HSC-like survey covering $676\ \rm{deg}^2$ that detects all satellites (i.e., systems with $M_V<0\ \rm{mag}$ and $r_{1/2}>10\ \rm{pc}$) down to a surface brightness of $\mu_V = 32\ \rm{mag}\ \rm{arcsec}^{-2}$ and out to a heliocentric distance of $300\ \rm{kpc}$, assuming an isotropic satellite distribution at high Galactic latitudes and accounting for subhalo disruption. There are six known satellites in the HSC footprint, but two of the six (Sextans and Leo IV) are detected at high significance in PS1 by at least one of the search algorithms in \citetalias{PaperI}. We find that our mock HSC survey detects~$1.75\pm 0.6$ satellites, which is in slight tension with the number of systems detected by HSC (four, after discounting Sextans and Leo IV).

\begin{deluxetable*}{{l@{\hspace{1.0in}}c@{\hspace{1.0in}}c}}[t]
\centering
\tablecolumns{3}
\tablecaption{Galaxy--Halo Connection Model Constraints Derived from Our Fit to the DES and PS1 Satellite Populations}
\tablehead{\colhead{Parameter\phantom{textttt}}{\hspace{1.0in}} & \colhead{Physical Interpretation\phantom{Interpretation text}} {\hspace{1.0in}}& \colhead{$95\%$ Confidence Interval}}
\startdata
Faint-end Slope
& Power-law slope of satellite luminosity function
& $-1.46 < \alpha < -1.39 $\\
\hline
Luminosity scatter
& Scatter in luminosity at fixed $V_{\rm{peak}}$
& $0\ \mathrm{dex}^{*} < \sigma_M < 0.19\ \rm{dex}$\\
\hline
$50\%$ occupation mass
& Mass at which $50\%$ of halos host galaxies
& $7.5^{*} < \log(\mathcal{M}_{50}/M_{\rm{\odot}}) < 7.93$\\
\hline
Baryonic effects
& Strength of subhalo disruption due to baryons
& $0.3 < \mathcal{B} < 2.1$\\
\hline
Occupation scatter
& Scatter in galaxy occupation fraction
& $0\ \rm{dex}^{*} < \sigma_{\rm{gal}} < 0.67\ \rm{dex}$\\
\hline
Size amplitude
& Amplitude of galaxy--halo size relation
& $0^{*}\ \rm{pc} < \mathcal{A} < 110\ \rm{pc}$\\
\hline
Size scatter
& Scatter in half-light radius at fixed halo size
& $0\ \mathrm{dex}^{*} < \sigma_{\log R} < 1.2\ \rm{dex}$\\
\hline
Size power-law index
& Power-law index of galaxy--halo size relation
& $0^{*} < n < 1.8$\\
\hline
\enddata
{\footnotesize \tablecomments{Asterisks mark prior-driven constraints.}}
\label{tab:constraints}
\end{deluxetable*}

Figure \ref{fig:tot_LFs} illustrates several predictions from hydrodynamic simulations of isolated and satellite dwarf galaxies. Our results are largely consistent with the luminosity function of bright MW satellites in hydrodynamic simulations of the Local Group using the Feedback In Realistic Environments (FIRE) feedback prescription, down to the FIRE resolution limit of $\sim -6\ \rm{mag}$ \citep{Garrison-Kimmel180604143}. Note that these FIRE simulations do not include LMC or SMC analogs, which accounts for the discrepancy with both our predictions and the observed luminosity function at $M_V<-16\ \rm{mag}$. Interestingly, other recent hydrodynamic simulations indicate that different star formation prescriptions significantly impact the amplitude and faint-end slope of the luminosity function for satellites of isolated LMC-like halos \citep{Munshi181012417}. Thus, our constraints on the faint-end slope, which are driven by satellites with $M_V>\ \sim -6\ \rm{mag}$ (corresponding to stellar mass $M_*<\ \sim 10^{5}\ M_{\rm{\odot}}$), can be used to inform subgrid star formation prescriptions.

\subsection{Galaxy--Halo Connection Model Constraints}
\label{constraints}

\begin{figure*}[t]
    \includegraphics[scale=0.41]{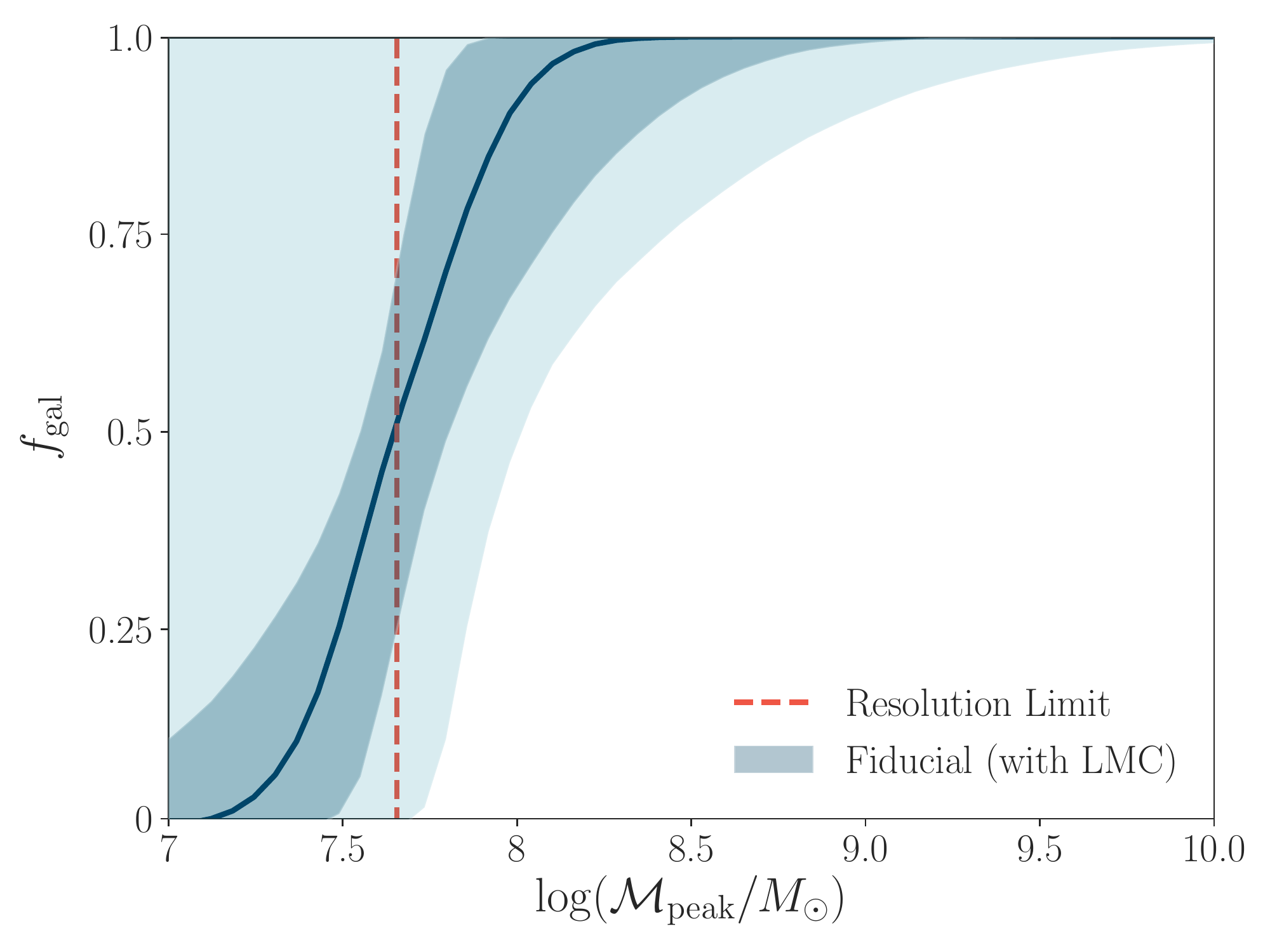}
    \includegraphics[scale=0.455]{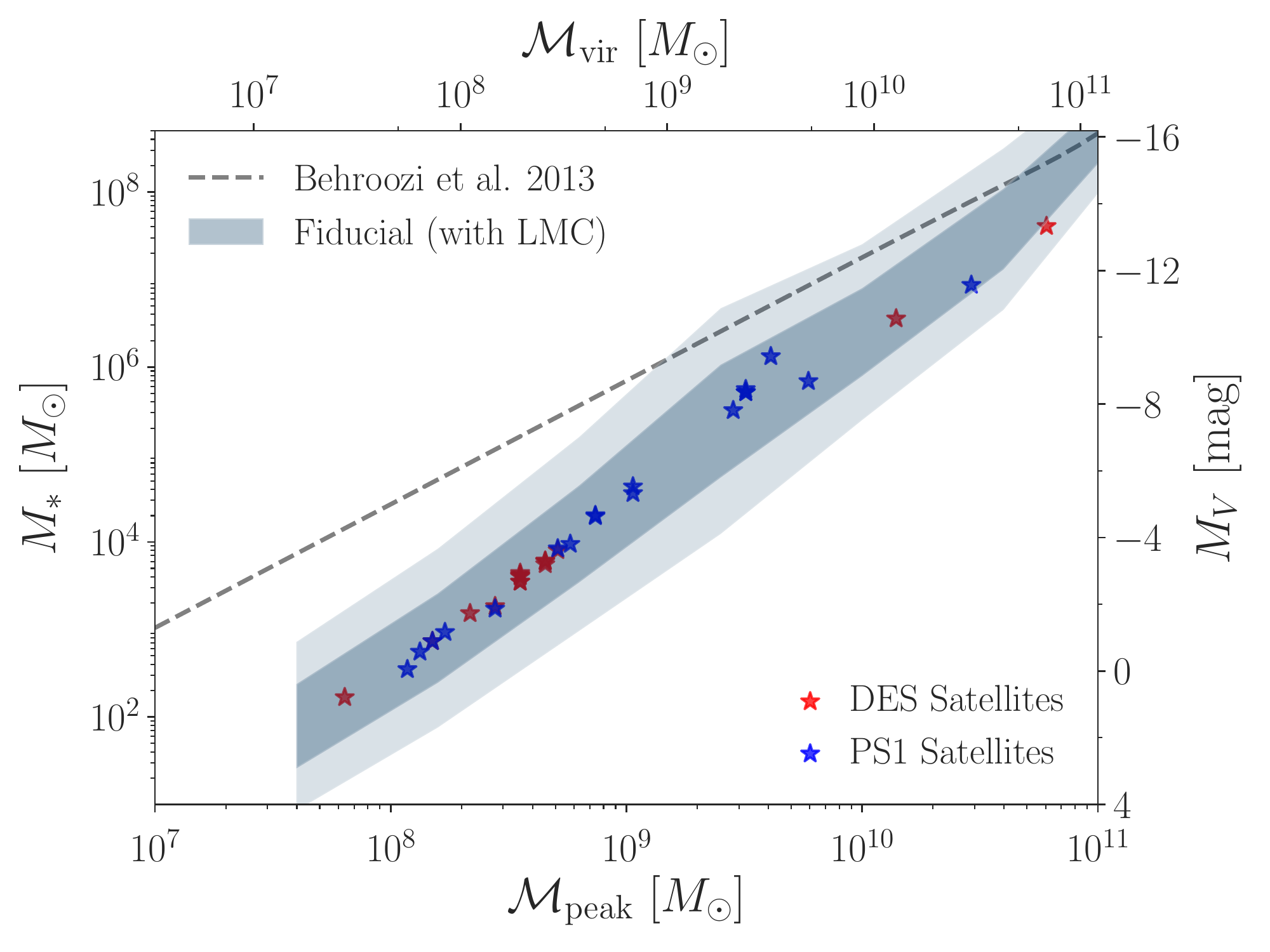}
    \caption{Left panel: fraction of halos that host galaxies, inferred from our fit to the DES and PS1 satellite populations. The solid line shows the median inferred galaxy occupation fraction, and dark (light) shaded contours represent $68\%$ ($95\%$) confidence intervals. The resolution limit of our simulations is indicated by the dashed vertical line. Right panel: SMHM relation inferred from our fit to the DES and PS1 satellite populations. An extrapolation of the mean SMHM relation derived from more luminous field galaxies is shown in gray \citep{Behroozi12076105}. Stars illustrate the mean of the predicted $\mathcal{M}_{\rm{peak}}$ range corresponding each observed DES and PS1 satellite, and top ticks indicate the corresponding present-day virial masses of the halos that host these systems.}
    \label{fig:fgal}
\end{figure*}

The posterior distribution for our fiducial model is shown in Figure \ref{fig:posterior}, and the corresponding galaxy--halo connection model constraints are listed in Table \ref{tab:constraints}. Note that we obtain statistically consistent results when fitting the DES and PS1 satellite populations with either of our two fiducial simulations individually in terms of both the Bayesian evidence and the galaxy--halo connection model constraints. In particular, no parameter constraints shift by more than $1\sigma$ relative to the fiducial values reported below when the fit is performed using either simulation individually. We now discuss each constraint in detail.
\begin{enumerate}
\item The inferred faint-end slope of the satellite luminosity function is steeper than that reported in a previous study based on classical and SDSS satellites \citep{Nadler180905542}. Our constraint is consistent with the faint-end slope derived from higher-luminosity field galaxies in the GAMA survey \citep{Loveday150501003,Wright170504074}, even though it is based on a sample that extends nearly $10\ \rm{mag}$ fainter than that used in GAMA. We note that~$\alpha$ is the most sensitive parameter in our analysis to modeling assumptions and details of the observed satellite population.
\item The scatter in luminosity at fixed $V_{\rm{peak}}$ is constrained to $\sigma_M < 0.19\ \rm{dex}$ at $95\%$ confidence, which may inform hydrodynamic feedback prescriptions that predict a steep increase in luminosity scatter at low masses (e.g., see \citealt{Wechsler180403097}). Our lack of a lower limit on $\sigma_M$ is consistent with previous studies of faint galaxy samples (e.g., \citealt{Lehmann151005651}). Meanwhile, large values of $\sigma_M$ are not allowed because too many low-$V_{\rm{peak}}$ satellites upscatter to observable luminosities, resulting in overpredicted luminosity functions. To confirm that this upper limit is robust, we calculate Bayes factors by drawing samples from the posterior in bins of~$\sigma_M$, finding that $\sigma_M = 0.15\ \rm{dex}$~($\sigma_M = 0.2\ \rm{dex}$) is disfavored relative to $\sigma_M=0\ \rm{dex}$ with a Bayes factor of $30$ ($100$).\footnote{We provide details on our Bayes factor calculations in Appendix \ref{appendixb3}.} These upper limits are comparable to the scatter typically inferred from abundance-matching analyses of brighter systems ($\sigma_M\sim 0.2\ \rm{dex}$) and smaller than that from hydrodynamic simulations of dwarf galaxies (e.g., \citealt{Rey190904664}); however, we caution that our constraint might be impacted by the use of only two independent realizations of the MW satellite population. In addition, it is potentially misleading to compare global constraints on scatter to those derived from the MW alone. Both of these caveats are important to explore in future work.
\item The peak mass at which $50\%$ of halos host galaxies is inferred to be less than $8.5\times10^7\ M_{\rm{\odot}}$ at $95\%$ confidence. Note that this summary statistic depends on the lower limit of our prior on $\mathcal{M}_{50}$, since the $\mathcal{M}_{50}$ posterior flattens near its lower limit, which is chosen based on the resolution of our simulations. Thus, we also calculate Bayes factors by drawing from the posterior in bins of $\mathcal{M}_{50}$ to confirm that this summary statistic is robust. We find that~$\mathcal{M}_{50}=8.5\times10^7\ M_{\rm{\odot}}$ ($\mathcal{M}_{50}=1.5\times 10^8\ M_{\rm{\odot}}$) is disfavored relative to arbitrarily low values of $\mathcal{M}_{50}$ with a Bayes factor of~$50$ ($100$). The current data are not able to place a lower limit on $\mathcal{M}_{50}$, which would correspond to the detection of a cutoff in galaxy formation.
\item Our posterior is consistent with $\mathcal{B}=1$, corresponding to our fiducial baryonic disruption model. Although a large spread in disruption strength is allowed by the data, extremely efficient ($\mathcal{B}>2.1$) and inefficient~($\mathcal{B}<0.3$) subhalo disruption relative to hydrodynamic simulations is strongly disfavored. These constraints widen when our lognormal prior on $\mathcal{B}$ is relaxed; however, zero subhalo disruption (corresponding to $\mathcal{B}=0$) is robustly ruled out.
\item The scatter in the galaxy occupation fraction is consistent with zero, which makes sense given our lack of a lower limit on $\mathcal{M}_{50}$. Models with large scatter in the occupation fraction ($\sigma_{\rm{gal}}<0.67\ \rm{dex}$), corresponding to extremely stochastic galaxy formation, are disfavored relative to a step function occupation fraction at $95\%$ confidence. Note that the slope of the~$\sigma_{\rm{gal}}$ posterior is driven by the lower limit of our $\mathcal{M}_{50}$ prior; as this limit decreases, the $\sigma_{\rm{gal}}$ posterior flattens.
\item The amplitude of the galaxy--halo size relation, defined as the typical size of a satellite in a halo with $\rvir=10\ \rm{kpc}$ at accretion, is constrained to lie between~$12$ and $73\ \rm{pc}$ at $68\%$ confidence. For larger values of $\mathcal{A}$, satellites are too large to be detected with high probability, and the DES and PS1 luminosity functions are underpredicted; for smaller values of~$\mathcal{A}$, many predicted satellites do not pass our $r_{1/2}>10\ \rm{pc}$ cut, and the luminosity functions are underpredicted.
\item The scatter in the galaxy--halo size relation is constrained to lie between $0.33$ and $0.91\ \mathrm{dex}$ at $68\%$ confidence. For larger values of $\sigma_{\log R}$, faint satellites upscatter to large sizes too frequently, which results in underpredicted luminosity functions. Our $68\%$ confidence lower limit on $\sigma_{\log R}$ of~$0.33\ \rm{dex}$ is consistent with the value estimated in~\cite{Kravstov12122980}. Lower values of $\sigma_{\log R}$ lead to slightly too many predicted DES and PS1 satellites; however, our results are consistent with $\sigma_{\log R}=0\ \mathrm{dex}$ at $95\%$ confidence.
\item The power-law index of the galaxy--halo size relation is constrained to lie between $0.5$ and $1.45$ at $68\%$ confidence. For shallower power-law slopes, satellite sizes do not change appreciably as a function of halo size, which results in a worse joint fit to the observed absolute magnitude and surface brightness distribution. We note that the posterior widens when our Gaussian prior on $n$ is relaxed.
\end{enumerate}

\subsection{Properties of Halos that Host the Faintest Satellites}
\label{faint}

We now explore the properties of the lowest-mass halos inferred to host MW satellites. The left panel of Figure \ref{fig:fgal} shows the galaxy occupation fraction derived from our statistical inference, where uncertainties are estimated by drawing from our fiducial posterior. By sampling from our fiducial posterior, we infer that halos with a peak virial mass below~$2.5\times 10^{8}\ M_{\rm{\odot}}$ and peak circular velocity below $19\ \rm{km\ s}^{-1}$ host at least one of the faintest observed satellites. To convert these into maximally conservative upper limits, we account for the uncertainty in MW host halo mass using the procedure described in Appendix \ref{appendixa1}, which yields limits on the minimum halo mass and peak circular velocity of~$\mathcal{M}_{\rm{min}}<3.2\times 10^{8}\ M_{\rm{\odot}}$ and $V_{\rm{peak,min}}<21\ \rm{km\ s}^{-1}$ at $95\%$ confidence. Furthermore, we predict that the faintest observed satellite inhabits a halo with $\mathcal{M}_{\rm{peak}}=1.5\times 10^{8}\ M_{\rm{\odot}}$, on average.\footnote{The faintest observed satellite in our analysis, Cetus II, is detected by DES with $M_V = 0.02\ \rm{mag}$ (\citealt*{Drlica-Wagner150803622}; Table \ref{fig:des_table}).}

These results improve upon the minimum halo mass constraint derived from classical and SDSS satellites \citep{Nadler180905542} by a factor of $2$, and they are consistent with the constraints reported in \cite{Jethwa161207834}. Moreover, these upper limits are not in significant tension with the expected atomic cooling limit of~$V_{\rm{peak}}\approx 20\ \rm{km\ s}^{-1}$, contrary to recent studies based on the radial MW satellite distribution (e.g., \citealt{Graus180803654}) and consistent with the findings in \cite{Bose190904039}.

We caution that the median galaxy occupation fraction shown in Figure \ref{fig:fgal} is driven by the assumed functional form in Equation \ref{eq:fgal} and is therefore arbitrary. Although the functional form in Equation \ref{eq:fgal} is consistent with results from hydrodynamic simulations for $\mathcal{M}_{\rm{peak}} \gtrsim 10^{9}\ M_{\rm{\odot}}$, this particular functional form is not required to fit the DES and PS1 luminosity functions. Rather, we have evidence that $f_{\rm{gal}}>50\%$ above a peak virial mass of $\sim 10^{8}\ M_{\rm{\odot}}$. To verify that the assumed form of the galaxy occupation fraction does not impact our constraints, we also test a binned model in which we fit for $\mathcal{M}_{50}$ and a corresponding $90\%$ occupation mass. We find that the resulting $50\%$ and $90\%$ occupation constraints are consistent with those inferred from our fiducial analysis.

A wide range of galaxy occupation fractions have been reported in hydrodynamic simulations, with some placing $\mathcal{M}_{50}$ as high as $\sim 10^9\ M_{\rm{\odot}}$ \citep{Sawala14066362,Fitts180106187}. However, recent hydrodynamic simulations run at higher resolution result in efficient galaxy formation in significantly lower-mass halos, and some claim that \emph{all} halos down to the resolution limit consistently host star particles \citep{Wheeler181202749}. In addition, simulations of galaxy formation at early pre-reionization epochs show that stellar systems form in halos with masses as low as $\sim 10^7\ M_{\rm{\odot}}$ (e.g., see Fig.\ 13 in \citealt{Cote171006442} for a compilation of recent simulation results). Most recently, high-resolution simulations of high-redshift galaxy formation that include the effects of spatially and temporally inhomogeneous reionization find $\mathcal{M}_{50}\sim 10^{8}\ M_{\rm{\odot}}$ \citep{Katz190511414}.

Our galaxy occupation fraction constraint implies that models with $\mathcal{M}_{50}>10^{8}\ M_{\rm{\odot}}$ are in significant tension with the observed MW satellite population, as long as MW satellite formation is representative of galaxy formation at this halo mass scale, on average. This assumption may not be true if the reionization history of the MW's Lagrangian volume differs from the average reionization history of an MW-mass halo hosting dwarf galaxies of the masses considered here (however, see \citealt{Alvarez08123405,Busha09013553}). Note that analyses based on H I surveys of Local Group dwarfs indicate a suppression mass scale similar to our $\mathcal{M}_{50}$ constraint \citep{Tollerud171100485}.

Due to our abundance-matching assumption, the lowest-mass halos in our model host the faintest galaxies, on average. Thus, our constraints on the masses of these halos are conservative, since the most massive halos in our simulations are forced to host more easily observable satellites at fixed distance and size, modulo baryonic disruption effects and abundance-matching scatter. In other words, our abundance-matching model yields a testable prediction: the faintest galaxies should inhabit the halos with the lowest pre-infall virial masses. We expect this correlation to be weakened by post-infall effects, including tidal stripping, but we can nevertheless infer the present-day joint distribution of halo mass and satellite luminosity or stellar mass. We illustrate this stellar mass--halo mass (SMHM) relation in the right panel of Figure \ref{fig:fgal}. Our inferred SMHM relation is generally consistent with recent results (e.g., \citealt{Jethwa161207834}). Like the faint-end slope of the luminosity function, the SMHM relation can be used to discriminate between different subgrid models of star formation and stellar feedback \citep{Munshi181012417}. As in previous studies, we find that the SMHM relation in the ultrafaint regime falls off more steeply than extrapolations of abundance-matching relations derived using higher-mass field galaxies \citep{Behroozi12076105}. Interestingly, \cite{Agertz190402723} found that full on-the-fly radiative transfer is necessary to match the steepness and normalization of our inferred SMHM relation for a fixed hydrodynamic feedback prescription.

Ultimately, our predictions must be confronted with the dynamical mass function of observed satellites, measurements of which will improve significantly in the era of upcoming spectroscopic facilities and giant segmented mirror telescopes \citep{Simon190304743}. A preliminary comparison of our joint predicted distribution of luminosity and $V_{\rm{max}}$ with the measured stellar velocity dispersions of DES and PS1 satellites suggests that our model is consistent with the inferred central densities of low-luminosity satellites~($M_V>-6\ \rm{mag}$). Although there is a systematic discrepancy between observed and predicted values of $V_{\rm{max}}$ for brighter systems (the ``too big to fail'' problem), our simple comparison does not account for the conversion from line-of-sight velocity dispersion measured within observed half-light radii to $V_{\rm{max}}$ or the tidal effects of the Galactic disk on the density profiles of surviving subhalos. Moreover, the systems for which predicted densities are higher than those inferred observationally are susceptible to baryonic feedback processes that core the inner regions of halos \citep{DiCinto13060898}, and this effect has been shown to alleviate the too big to fail problem \citep{Brooks12095394,Sawala151101098,Wetzel160205957,Lovell161100005,Garrison-Kimmel180604143}.

Finally, we explore the properties of the halos inferred to host the faintest potentially detectable galaxies. In particular, we calculate the minimum peak halo mass necessary for halos to contain a stellar population of at least $100\ M_{\rm{\odot}}$, chosen to represent the approximate threshold for which it would be possible to observationally confirm a stellar system as a dark matter--dominated dwarf galaxy.\footnote{For many MW satellites, this will likely require spectroscopy with giant segmented mirror telescopes \citep{Simon190304743}.} By populating a higher-resolution version of one of our fiducial simulations and sampling from the posterior of our abundance-matching relation, we find that systems at the observational threshold occupy halos with~$\mathcal{M}_{\rm{peak}} > 10^6\ M_{\rm{\odot}}$ at $95\%$ confidence. To detect even lower-mass halos, gravitational probes of dark matter that are independent of baryonic content, e.g.\ gravitational lensing or stellar streams, must be employed.

\subsection{Implications for Dark Matter Microphysics}
\label{microphysics}

\begin{figure}[t]
    \includegraphics[scale=0.385]{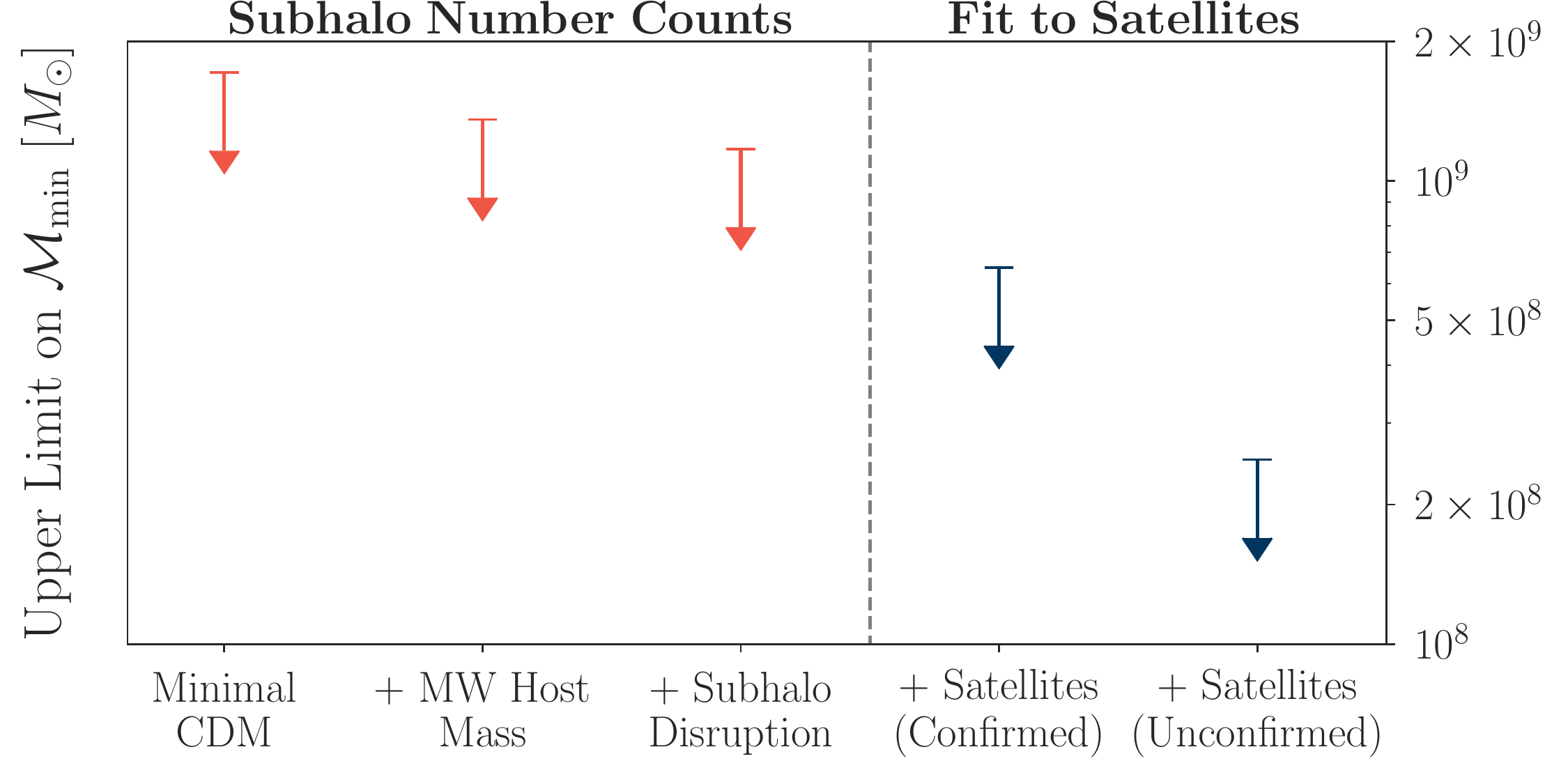}
    \caption{Impact of modeling assumptions on the minimum subhalo mass inferred from the observed DES and PS1 satellite populations. The first three models match the number of subhalos to the number of confirmed DES and PS1 satellites, and the last two models populate subhalos with galaxies to fit the position-dependent MW satellite luminosity function and size distribution.}
    \label{fig:uncertainties}
\end{figure}

Many deviations from CDM lead to a cutoff in the abundance of low-mass halos. Several authors have used MW satellite abundances to constrain a free-streaming cutoff induced by warm dark matter (e.g., \citealt{Maccio09102460,Kennedy14,Lovell13081399,Jethwa161207834}). \cite{Nadler190410000} showed that similar constraints apply to other dark matter models, resulting in limits on the velocity-independent scattering cross section between dark matter and baryons. Our statistical detection of halos with peak virial masses below $3.2\times 10^{8}\ M_{\rm{\odot}}$ therefore translates directly into constraints on various microphysical properties of dark matter.

\cite{Nadler190410000} found that the minimum halo mass inferred in this fashion is comparable to the limit on the half-mode mass $M_{\rm{hm}}$, which corresponds to the scale at which the matter power spectrum is suppressed by a critical amount relative to CDM due to dark matter free streaming or interactions. Performing a statistical inference in which the half-mode mass is varied will constrain it to lie below our upper limit on $\mathcal{M}_{50}$, since the abundance of halos at and above the half-mode mass is reduced relative to CDM.

Thus, for a simple and conservative estimate of the dark matter constraints resulting from our analysis, we set the upper limit on $M_{\rm{hm}}$ equal to our upper limit on the minimum halo mass, i.e., $M_{\rm{hm}}<3.2\times 10^{8}\ M_{\rm{\odot}}$, which corresponds to a lower limit on the half-mode scale of $k_{\rm{hm}}>36\ h\ \rm{Mpc}^{-1}$. Using the relations in \cite{Nadler190410000} with the cosmological parameters $h = 0.7$ and $\Omega_{\rm m} = 0.286$ corresponding to the simulations used in our analysis (see Section \ref{description}), this yields a lower limit of $3.4\ \rm{keV}$ on the mass of thermal relic warm dark matter and an upper limit of~$6\times 10^{-30}\ \rm{cm}^2$ on the velocity-independent dark matter--baryon scattering cross section for a $10\ \rm{keV}$ dark matter particle mass, both at $95\%$ confidence. We leave a detailed investigation of dark matter constraints to future work.


\section{Theoretical Uncertainties}
\label{uncertainties}

We aim to present a thorough galaxy--halo connection model that allows us to marginalize over the most important theoretical uncertainties when modeling the MW satellite population. Nonetheless, our modeling choices necessarily affect the predicted number of detected low-mass halos and thus our upper limit on the minimum halo mass, $\mathcal{M}_{\mathrm{min}}$. In this section, we briefly discuss the main uncertainties in this analysis and their impact on our $\mathcal{M}_{\mathrm{min}}$ constraint.

To do so, we consider the upper limit on $\mathcal{M}_{\mathrm{min}}$ as a function of modeling assumptions, starting with the most conservative model possible and adding one assumption at a time. We illustrate the results of this exercise in Figure \ref{fig:uncertainties} for upper limits calculated as follows.
\begin{enumerate}
\item Minimal CDM. Assuming a maximally massive MW halo given Gaia constraints~(i.e., a virial mass of~$1.8\times 10^{12}\ M_{\rm{\odot}}$; \citealt{Callingham180810456,Cautun191104557,Li191011257,Li191202086}), count the subhalos within the virial radius of the MW in order of decreasing $V_{\rm{peak}}$ until the number of kinematically confirmed DES and PS1 satellites is matched, and set the lowest corresponding value of $\mathcal{M}_{\rm{peak}}$ equal to the upper limit on $\mathcal{M}_{\mathrm{min}}$.
\item MW Host Mass. Repeat the previous step with the MW host mass fixed to its average value in our two fiducial simulations~(i.e., an average virial mass of~$1.4\times 10^{12}\ M_{\rm{\odot}}$).
\item Subhalo Disruption. Repeat the previous step many times with the subhalo number weighted by disruption probability, sampling $\mathcal{B}$ from our fiducial posterior, to calculate an upper limit on $\mathcal{M}_{\mathrm{min}}$ at $95\%$ confidence.
\item Satellites (confirmed). Repeat the previous step including the observational detection probabilities for mock satellites in the DES and PS1 footprints by drawing satellite properties from our fiducial posterior.
\item Satellites (unconfirmed). Repeat the previous step including the unconfirmed candidate satellites detected by DES and PS1 in the observed tally.
\end{enumerate}
This yields~$\mathcal{M}_{\mathrm{min}} <(17, 14, 12, 6.5, 2.5)\times 10^{8}\ M_{\rm{\odot}}$ for models (i)--(v), respectively. Note that models (i)--(iii) are extremely conservative, since subhalos are counted in order of decreasing $V_{\rm{peak}}$; however, these models do not reproduce the observed position-dependent MW satellite luminosity function or radial distribution. Model (iv) yields the conservative limit presented in Appendix \ref{appendixd1}, and model (v) yields our fiducial constraint, uncorrected for MW host halo mass. Although we have not explicitly considered artificial subhalo disruption in this list of theoretical uncertainties (e.g., \citealt{VandenBosch180105427,VandenBosch171105276}), our fiducial orphan satellite model effectively assumes that subhalo disruption in dark matter--only simulations is entirely artificial, which is a conservative choice.

Figure \ref{fig:uncertainties} shows that both fitting the satellite luminosity function and including the population of faint, kinematically unconfirmed satellite galaxies in our fit yield significant increases in constraining power. We emphasize that our galaxy--halo connection model is conservative from the perspective of upper limits on the minimum halo mass, since we assume that high-mass halos host the brightest observed satellite galaxies. Moreover, we marginalize over many uncertainties in the connection between low-mass halos and faint galaxies. Thus, the largest gain in constraining power likely results from our detailed use of observational selection functions, i.e., from the fact that some satellites are not detected in DES or PS1 data. Given our extensive validation of the DES and PS1 selection functions in \citetalias{PaperI}, we are therefore confident in our minimum halo mass constraints.


\section{Conclusions}
\label{discussion}

We have presented the results of a forward-modeling framework for MW satellites applied to recent searches for satellites in photometric surveys over nearly the entire high Galactic latitude sky. Our analysis includes position-dependent observational selection effects that faithfully represent satellite searches in DES and PS1 imaging data, and our galaxy--halo connection model allows us to marginalize over theoretical uncertainties in the relationship between galaxy and halo properties, the effects of baryonic physics on subhalo populations, and the stochastic nature of galaxy formation in low-mass halos. By performing a Bayesian analysis of the observed DES and PS1 satellite populations, we find decisive statistical evidence for the following.
\begin{enumerate}
\item The LMC impacts the observed MW satellite population, contributing $4.8\pm 1.7$ ($1.1\pm 0.9$) LMC-associated satellites to the DES (PS1) satellite populations.
\item The LMC fell into the MW within the last $2\ \rm{Gyr}$.
\item The faintest satellites currently known occupy halos with peak virial masses less than $3.2\times 10^{8}\ M_\odot$.
\item The faintest detectable satellites (i.e., dark matter--dominated systems with $M_*>100\ M_{\rm{\odot}}$) occupy halos with peak virial masses greater than $10^{6}\ M_\odot$.
\end{enumerate}

These results have broad implications for galaxy formation and dark matter physics. For example, comparing our inferred luminosity function and galaxy occupation fraction to predictions from hydrodynamic simulations will help break degeneracies among subgrid star formation and feedback models. Meanwhile, extending our model to study the evolution of the luminosity function will shed light on high-redshift faint galaxy populations (e.g., \citealt{BoylanKolchin150406621,Weisz170206129}) and the MW's local reionization history (e.g., \citealt{Busha09013553,Lunnan11052293,Katz190511414}).

Finally, our statistical detection of low-mass halos translates directly into constraints on a suite of dark matter properties, including warmth in thermal production scenarios, initial velocity distribution in nonthermal production scenarios, self-interaction cross section, interaction strength with the Standard Model, formation redshift, stability, and quantum mechanical behavior on astrophysical scales. Exploring the interplay between galaxy formation physics and alterations to the standard CDM paradigm will be crucial in order to extract these signals from upcoming observations of ultrafaint galaxies, and forward-modeling approaches like the one developed here will drive these studies forward.


\acknowledgments

This paper has gone through internal review by the DES collaboration. Our code and subhalo catalogs are available at \href{https://github.com/eonadler/subhalo_satellite_connection}{github.com/eonadler/subhalo\_satellite\_connection}. We thank Ralf Kaehler for providing the  visualizations of our simulations used in Figure \ref{fig:framework}. We thank Shea Garrison-Kimmel and Andrew Wetzel for sharing data from the ELVIS on FIRE simulations. We thank Susmita Adhikari, Arka Banerjee, Ekta Patel, and Andrew Pontzen for useful conversations. Finally, we are grateful to the anonymous referee for constructive feedback.

This research received support from the National Science Foundation (NSF) under grant No.\ NSF AST-1517422, grant No.\ NSF PHY17-48958 through the Kavli Institute for
Theoretical Physics program ``The Small-Scale Structure of Cold(?)\ Dark Matter,''
and grant No.\ NSF DGE-1656518 through the NSF Graduate Research Fellowship received by E.O.N. 
Support for Y.-Y.M.\ was provided by the Pittsburgh Particle Physics, Astrophysics and Cosmology Center through the Samuel P.\ Langley PITT PACC Postdoctoral Fellowship and by NASA through the NASA Hubble Fellowship grant No.\ HST-HF2-51441.001 awarded by the Space Telescope Science Institute, which is operated by the Association of Universities for Research in Astronomy, Incorporated, under NASA contract NAS5-26555.
Part of this work was performed at the Aspen Center for Physics, which is supported by National Science Foundation grant PHY-1607611.

This research made use of computational resources at SLAC National Accelerator Laboratory, a U.S.\ Department of Energy Office; the authors are thankful for the support of the SLAC computational team. 
This research made use of the Sherlock
cluster at the Stanford Research Computing Center (SRCC); the authors are thankful for the support of the SRCC team. 
This research made use of \url{https://arXiv.org} and NASA's Astrophysics Data System for bibliographic information.

Funding for the DES Projects has been provided by the U.S. Department of Energy, the U.S. National Science Foundation, the Ministry of Science and Education of Spain, 
the Science and Technology Facilities Council of the United Kingdom, the Higher Education Funding Council for England, the National Center for Supercomputing 
Applications at the University of Illinois at Urbana-Champaign, the Kavli Institute of Cosmological Physics at the University of Chicago, 
the Center for Cosmology and Astro-Particle Physics at the Ohio State University,
the Mitchell Institute for Fundamental Physics and Astronomy at Texas A\&M University, Financiadora de Estudos e Projetos, 
Funda{\c c}{\~a}o Carlos Chagas Filho de Amparo {\`a} Pesquisa do Estado do Rio de Janeiro, Conselho Nacional de Desenvolvimento Cient{\'i}fico e Tecnol{\'o}gico and 
the Minist{\'e}rio da Ci{\^e}ncia, Tecnologia e Inova{\c c}{\~a}o, the Deutsche Forschungsgemeinschaft, and the Collaborating Institutions in the Dark Energy Survey. 

The Collaborating Institutions are Argonne National Laboratory, the University of California at Santa Cruz, the University of Cambridge, Centro de Investigaciones Energ{\'e}ticas, 
Medioambientales y Tecnol{\'o}gicas-Madrid, the University of Chicago, University College London, the DES-Brazil Consortium, the University of Edinburgh, 
the Eidgen{\"o}ssische Technische Hochschule (ETH) Z{\"u}rich, 
Fermi National Accelerator Laboratory, the University of Illinois at Urbana-Champaign, the Institut de Ci{\`e}ncies de l'Espai (IEEC/CSIC), 
the Institut de F{\'i}sica d'Altes Energies, Lawrence Berkeley National Laboratory, the Ludwig-Maximilians Universit{\"a}t M{\"u}nchen and the associated Excellence Cluster Universe, 
the University of Michigan, the NSF’s National Optical-Infrared Astronomy Laboratory, the University of Nottingham, The Ohio State University, the University of Pennsylvania, the University of Portsmouth, 
SLAC National Accelerator Laboratory, Stanford University, the University of Sussex, Texas A\&M University, and the OzDES Membership Consortium.

Based in part on observations at Cerro Tololo Inter-American Observatory, NSF’s National Optical-Infrared Astronomy Laboratory, which is operated by the Association of 
Universities for Research in Astronomy (AURA) under a cooperative agreement with the National Science Foundation.

The DES data management system is supported by the National Science Foundation under grant Nos.\ AST-1138766 and AST-1536171.
The DES participants from Spanish institutions are partially supported by MINECO under grants AYA2015-71825, ESP2015-66861, FPA2015-68048, SEV-2016-0588, SEV-2016-0597, and MDM-2015-0509, 
some of which include ERDF funds from the European Union. The IFAE is partially funded by the CERCA program of the Generalitat de Catalunya.
Research leading to these results has received funding from the European Research
Council under the European Union's Seventh Framework Program (FP7/2007-2013), including ERC grant agreements 240672, 291329, and 306478.
We  acknowledge support from the Brazilian Instituto Nacional de Ci\^encia
e Tecnologia (INCT) e-Universe (CNPq grant 465376/2014-2).

This manuscript has been authored by the Fermi Research Alliance, LLC, under contract No. DE-AC02-07CH11359 with the U.S. Department of Energy, Office of Science, Office of High Energy Physics.

\software{
Astropy \citep{astropy},
ChainConsumer \citep{ChainConsumer},
emcee \citep{emcee},
Healpy (\http{healpy.readthedocs.io}),
Pandas \citep{pandas},
\texttt{Scikit-Learn} \citep{scikit-learn}, 
IPython \citep{ipython},
Jupyter (\http{jupyter.org}),
Matplotlib \citep{matplotlib},
NumPy \citep{numpy},
SciPy \citep{scipy}, 
Seaborn (\https{seaborn.pydata.org}).
}

\bibliographystyle{yahapj2}
\bibliography{references,software}


\appendix

\section{Galaxy--Halo Connection Model Details}
\label{appendixa}

In this appendix, we examine several components of our galaxy--halo connection model to determine whether any of our assumptions significantly impact the results presented above.

\subsection{MW Host Halo Mass}
\label{appendixa1}

The analysis in this paper is restricted to two fiducial MW-like hosts with virial masses of $1.57$ and $1.26\times 10^{12}\ M_{\rm{\odot}}$. However, we expect the uncertainty in the MW host halo mass, $M_{\rm{MW}}$, to impact our constraints. Consider a toy model in which $N_{\rm{sat}}$ satellites brighter than a limiting magnitude~$M_{V,\rm{min}}$ must be predicted in order to match the observed luminosity function. In this toy model, the predicted number of satellites is given by
\begin{align}
    N_{\rm{sat}}(<M_{V,\rm{lim}}) &= \int_{M_{V,\rm{lim}}}^{-\infty} \frac{\mathrm{d}N_{\rm{sat}}}{\mathrm{d}M_V}\mathrm{d}M_V&\nonumber\\ &= \int_{\mathcal{M}_{\rm{min}}(M_{V,\rm{lim}})}^{\infty} f\Big(\frac{\mathrm{d}N_{\rm{sub}}}{\mathrm{d}\mathcal{M}_{\mathrm{peak}}},\boldsymbol{\theta}\Big)\mathrm{d}\mathcal{M}_{\mathrm{peak}},&
\end{align}
where $\mathrm{d}N_{\rm{sub}}/\mathrm{d}\mathcal{M}_{\mathrm{peak}}$ is the subhalo mass function, $\mathcal{M}_{\rm{min}}$ is the lowest halo mass populated by an observed satellite, and~$f$ encapsulates the observational selection, subhalo disruption, and galaxy occupation effects that determine whether each halo hosts an observable satellite, all of which depend on galaxy--halo connection model parameters $\boldsymbol{\theta}$. Neglecting the dependence of the latter effects on host mass (which we expect to be subdominant compared to the overall rescaling of subhalo abundances), using the standard linear relationship between subhalo abundance and host mass, and assuming a standard subhalo mass function slope of $\mathrm{d}N_{\rm{sub}}/\mathrm{d}\mathcal{M}_{\mathrm{peak}} \propto \mathcal{M}_{\mathrm{peak}}^{-2}$ (e.g., \citealt{Mao150302637}), we have
\begin{equation}
   N_{\rm{sat}}\propto M_{\rm{MW}}\int_{\mathcal{M}_{\rm{min}}}^{\infty} \mathcal{M}_{\mathrm{peak}}^{-2}\ \mathrm{d}\mathcal{M}_{\mathrm{peak}} \propto \frac{M_{\rm{MW}}}{\mathcal{M}_{\rm{min}}}.
\end{equation}
Thus, for a fixed observed satellite count $N_{\rm{sat}}$, we expect our~$95\%$ confidence level upper limit on $\mathcal{M}_{\mathrm{min}}$ to scale linearly with host mass. In addition, because the error on MW mass is independent of the error on $\mathcal{M}_{\mathrm{min}}$, we expect these uncertainties to add in quadrature.

Given our fiducial minimum halo mass of $2.5\times 10^{8}\ M_{\rm{\odot}}$ derived for an average host mass of $1.4\times 10^{12}\ M_{\rm{\odot}}$, we therefore expect $\mathcal{M}_{\mathrm{min}}<3.2\times 10^{8}\ M_{\rm{\odot}}$ ($\mathcal{M}_{\mathrm{min}}<2\times 10^{8}\ M_{\rm{\odot}}$) for a maximally high-mass (maximally low-mass) host halo given the current $2\sigma$ observational uncertainty on the MW virial mass of $1.0\times 10^{12}<M_{\rm{MW}}/M_{\rm{\odot}}<1.8\times 10^{12}$ \citep{Callingham180810456,Cautun191104557,Li191011257,Li191202086}. We expect the remaining galaxy--halo connection model parameters and associated errors to remain largely unchanged, although rerunning our analysis using additional simulations is required to confirm this hypothesis. We expect the inferred total satellite count to scale linearly with MW mass; thus, given our fiducial prediction of $220$ total MW satellites with~$M_V<0\ \rm{mag}$ and $r_{1/2}>10\ \rm{pc}$, we expect $280$ ($170$) such satellites for a maximally high-mass (maximally low-mass) host halo.

Above, we implicitly assumed that our galaxy--halo connection model is capable of adjusting the number of faint satellites, which make the largest contribution to $N_{\rm{sat}}$, while simultaneously matching the bright end of the observed satellite luminosity functions. This assumption holds because we have fixed the abundance-matching prescription to the relation derived from GAMA data for $M_V<-13\ \rm{mag}$ while allowing the faint-end slope to vary. We note that the results of \cite{NewtonMNRAS} suggest that the inferred number of satellites within a fixed physical radius is independent of $M_{\rm{MW}}$. We find that the total number of satellites inferred within the virial radius scales almost exactly linearly with~$M_{\rm{MW}}$, as expected from the linear scaling of subhalo abundance with host halo mass, and further confirming the consistency of the results among our two fiducial simulations. In addition, we find that $\mathcal{M}_{\mathrm{min}}$ is roughly independent of $M_{\rm{MW}}$ for our fiducial simulations.

\subsection{Mass-dependent Scatter} 
\label{appendixa2}

Here we test a model where the abundance-matching scatter in luminosity at fixed $V_{\rm{peak}}$,~$\sigma_M$, depends on peak halo mass. Motivated by the model in \cite{GarrisonKimmel160304855}, we set
\begin{equation}
    \sigma_M \equiv \sigma_{M,0} - \gamma_M(\log \mathcal{M}_{\rm{peak}}-\log \mathcal{M}_1),\label{eq:mass_scatter}
\end{equation}
where $\sigma_{M,0}$ is a free parameter that captures the amplitude of the luminosity scatter, $\gamma_M$ is a free parameter that captures its mass dependence, and $\mathcal{M}_1=10^{11}\ M_{\rm{\odot}}$ is fixed. By rerunning our fit with $\gamma_M$ as an additional ninth free parameter, we find that large values of $\gamma_M$ are ruled out by the DES and PS1 satellite populations at high statistical significance, with~$\gamma_M < 0.07$ at $95\%$ confidence. Large values of $\gamma_M$ are disfavored because abundant, low-mass halos host satellites that upscatter to observable luminosities too often to match the observed DES and PS1 luminosity functions; however, the same caveats noted in \S\ref{constraints} for our constraint on $\sigma_M$ apply to $\gamma_M$, so this upper limit should be interpreted with caution. Introducing mass-dependent scatter does not significantly affect our inferred upper bound on~$\mathcal{M}_{50}$, implying that our fiducial minimum halo mass constraint does not depend on the details of our luminosity scatter model.

\subsection{Radial Scaling} 
\label{appendixa3}

To account for potential biases in our radial subhalo distributions due to artificial disruption and halo finder incompleteness, we define the parameter $\chi$ by
\begin{equation}
r_{\rm{sat}} \equiv \chi r_{\rm{sub}},
\end{equation}
where $r_{\rm{sat}}$ is a satellite's distance from the center of its host halo, which we equate to its galactocentric distance, and $r_{\rm{sub}}$ is the galactocentric distance of the corresponding subhalo.

\begin{figure*}[t]
\centering
    \includegraphics[scale=0.35]{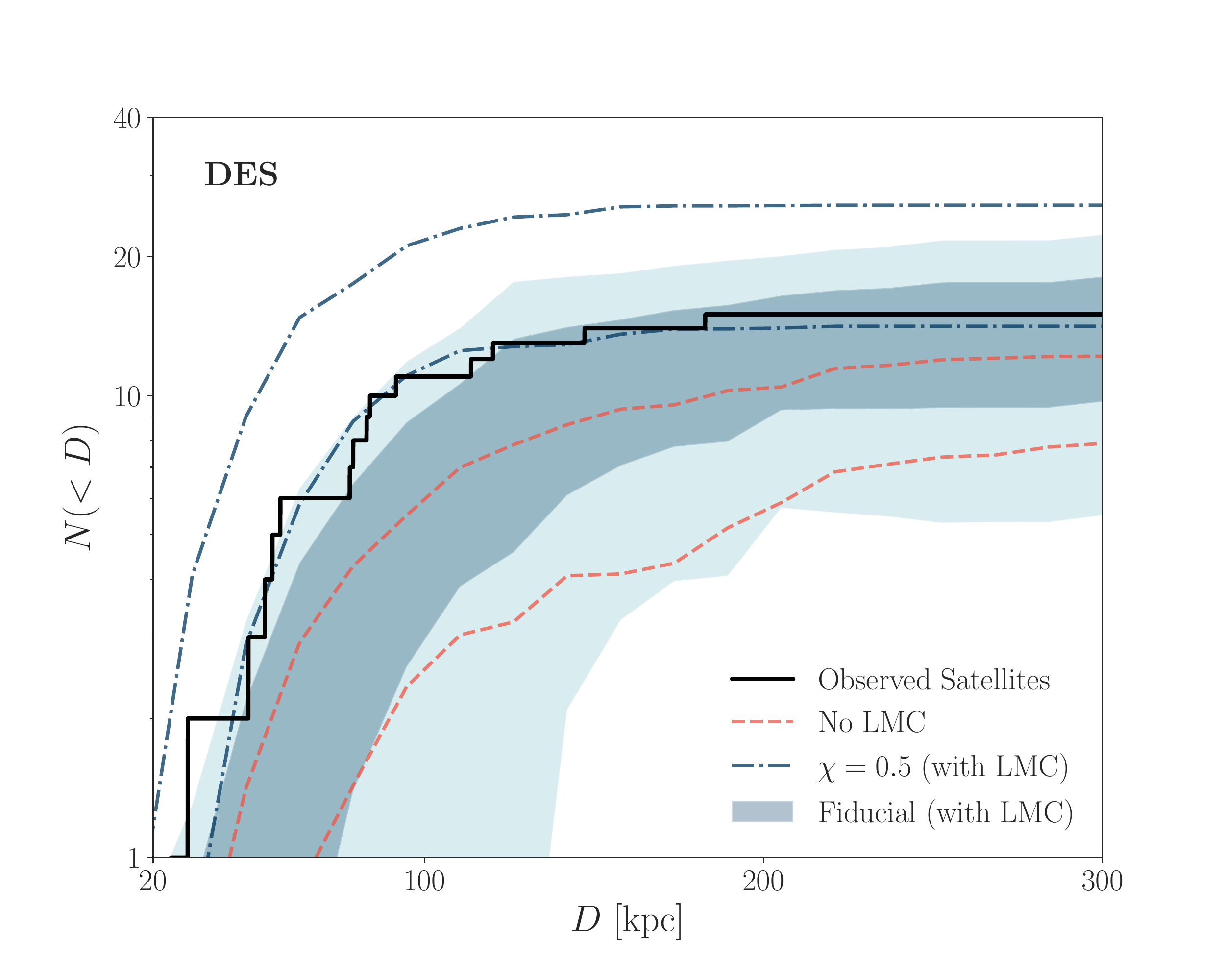}
    \includegraphics[scale=0.35]{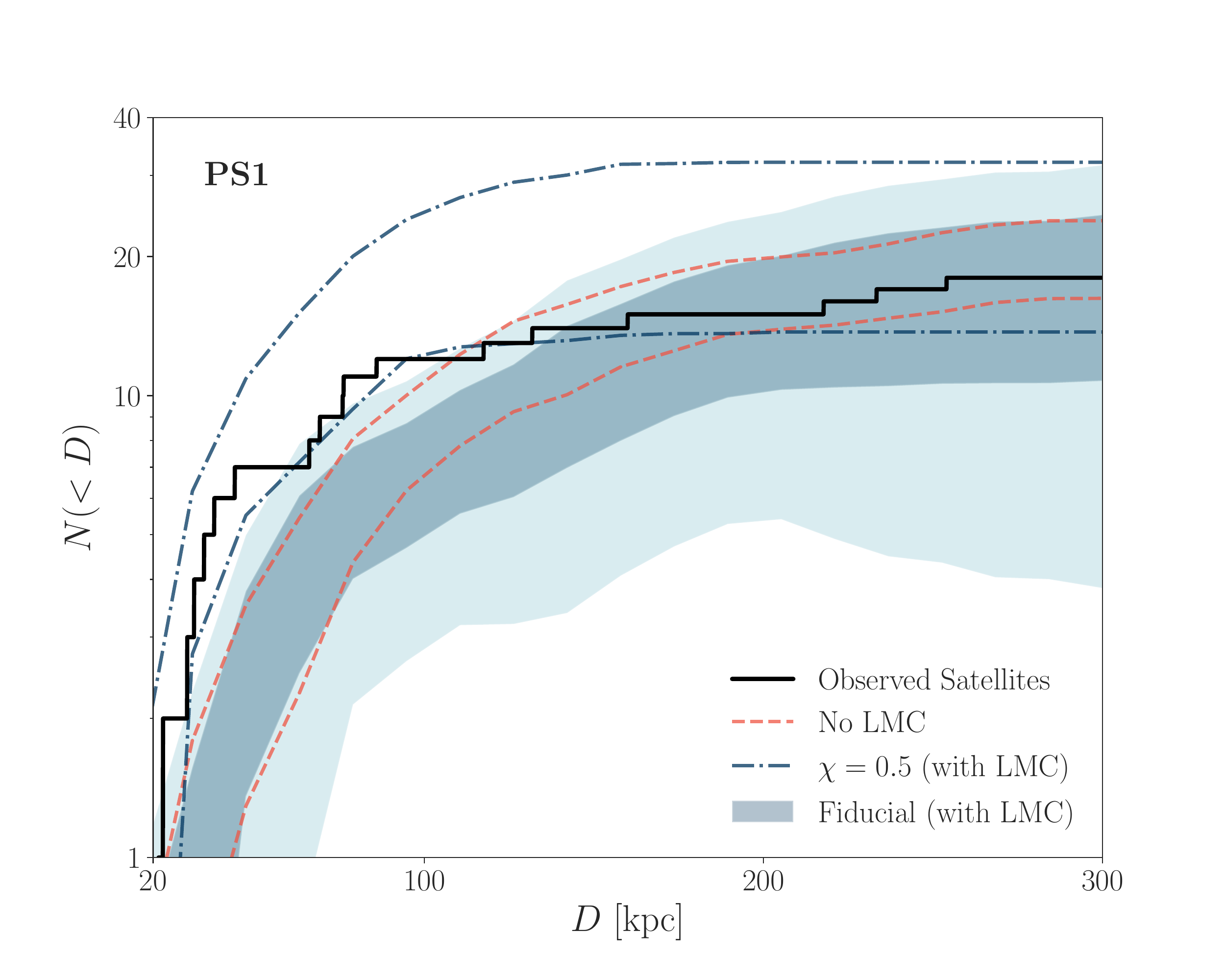}
    \caption{Radial distributions derived from our fit to the DES and PS1 satellite populations. Our fiducial eight-parameter galaxy occupation fraction model is shown in blue. Dark (light) blue bands correspond to $68\%$ ($95\%$) confidence intervals, dashed red lines show the $68\%$ confidence interval for a model using host halos without LMC analogs (``No LMC''), and black lines show the observed radial distributions. Dotted-dashed blue lines show the $68\%$ confidence interval for a model with a radial scaling parameter of $\chi=0.5$.}
    \label{fig:radial_dists}
\end{figure*}

In our main analysis, we take subhalo positions directly from the simulation data and therefore assume $\chi=1$. However, as noted above, our fiducial model slightly underpredicts the observed radial distribution of satellites close to the center of the MW in the PS1 footprint. We plot the predicted DES and PS1 radial distributions for our fiducial model in Figure~\ref{fig:radial_dists}; to illustrate the effect of varying $\chi$, we also show the $68\%$ confidence interval for our fiducial posterior evaluated with $\chi=0.5$.

To test the impact of radial scaling, we refit the DES and PS1 satellite populations with $\chi$ as an additional ninth free parameter. As expected, decreasing $\chi$ reduces the tension between the predicted and observed inner radial distribution of PS1 satellites; however, doing so does not significantly affect the goodness of fit for the observed luminosity functions and size distributions. Moreover, our key constraints, including the upper limit on $\mathcal{M}_{50}$, and our conclusions regarding the impact of the LMC system are not affected. In particular, the Bayes factors in favor of our fiducial LMC model relative to the alternative LMC scenarios defined in \S\ref{LMCimpact} are unchanged. We note that, since we have only fit to observed absolute magnitudes and surface brightnesses, the discrepancy with the observed radial distribution for $\chi=1$ might not persist for a fit that includes galactocentric distance; we comment on the technical difficulties associated with such a fit in Appendix \ref{appendixb4}.

\cite{Bose190904039} suggested that the Gaia-Enceladus accretion event, in which an LMC-mass galaxy merged with the MW $8$--$11\ \rm{Gyr}$ ago, might lead to a relative overabundance of ultrafaint satellites in the MW. Because of dynamical friction, this overabundance would be particularly evident in the innermost regions of the MW and might affect the observed radial satellite distribution. Interestingly, both host halos used in this work experience a Gaia-Enceladus-like accretion event, following the definition in \cite{Bose190904039} of a massive ($\sim 10^{11}\ M_{\rm{\odot}}$) halo merging with the MW halo between $z=1$ and $2$. Given that we still predict an underabundance of observed satellites in the inner regions of both fiducial host halos, the Gaia-Enceladus-like events they experience do not seem to be sufficient to ease the tension with the observed radial distribution. Nonetheless, exploring the relationship between the mass accretion history of the MW and the present-day radial distribution of observed ultrafaint satellites in detail is an interesting avenue for future work.

\subsection{Tidal Stripping} 
\label{appendixa4}

Following \cite{Nadler180905542}, we test a model for the evolution of satellite sizes by changing the mean sizes predicted by Equation \ref{eq:size} to
\begin{equation}
    r_{1/2}' \equiv r_{1/2}\ \left(\frac{V_{\rm{max}}}{V_{\rm{acc}}}\right)^{\beta},\label{eq:size_stripping} 
\end{equation}
where $r_{1/2}'$ denotes the satellite half-light radius at $z=0$, $r_{1/2}$ is the half-light radius at accretion predicted by Equation \ref{eq:size}, $V_{\rm{max}}$ ($V_{\rm{acc}}$) is the maximum circular velocity of a subhalo today (at accretion), and $\beta>0$ is a parameter that controls the strength of size reduction due to tidal stripping. We set $\beta=0$ in our fiducial analysis, meaning that satellite sizes are fixed based on halo sizes at accretion. However, tidal stripping after infall can shrink satellite sizes; for example, \cite{Penarrubia08111579} found that $1<\beta<2$ describes the results of high-resolution simulations well.

In Figure \ref{fig:tidal_stripping}, we illustrate predicted size distributions for our fiducial posterior evaluated with $\beta=3$; a large value of $\beta$ was chosen to test an extreme dependence of satellite sizes on tidal stripping. We find that even this extreme model does not impact the observed satellite size distributions, indicating that our results are robust to assumptions about tidal stripping. Our simulations lack the spatial resolution to test whether the \cite{Penarrubia08111579} prescription holds in detail and alters observed satellite size distributions, but this---along with an exploration of size enlargement due to tidal heating---is an interesting avenue for future work.

\subsection{Concentration-dependent Satellite Sizes}
\label{appendixa5}

\cite{Jiang180407306} found that galaxy sizes in two hydrodynamic simulations follow a size relation similar to that in \cite{Kravstov12122980}, with an additional dependence on halo concentration. In particular, the size relation
\begin{equation}
r_{1/2} \equiv \mathcal{A} \left ( \frac{c}{10} \right )^\gamma \left(\frac{R_{\rm{vir}}}{R_0}\right)^n,
\label{eqn:size}
\end{equation}
with $\mathcal{A}=0.02$, $n=1$, $\gamma=-0.7$, $R_0=1\ \rm{kpc}$, and halo concentration $c$ measured as a function of redshift, fits the hydrodynamic simulation results in \cite{Jiang180407306} with a residual scatter of $\sim 0.15\ \rm{dex}$. This relation implies that more concentrated halos host less extended stellar systems at a fixed virial radius in these simulations.

\begin{figure*}[]
    \includegraphics[scale=0.35]{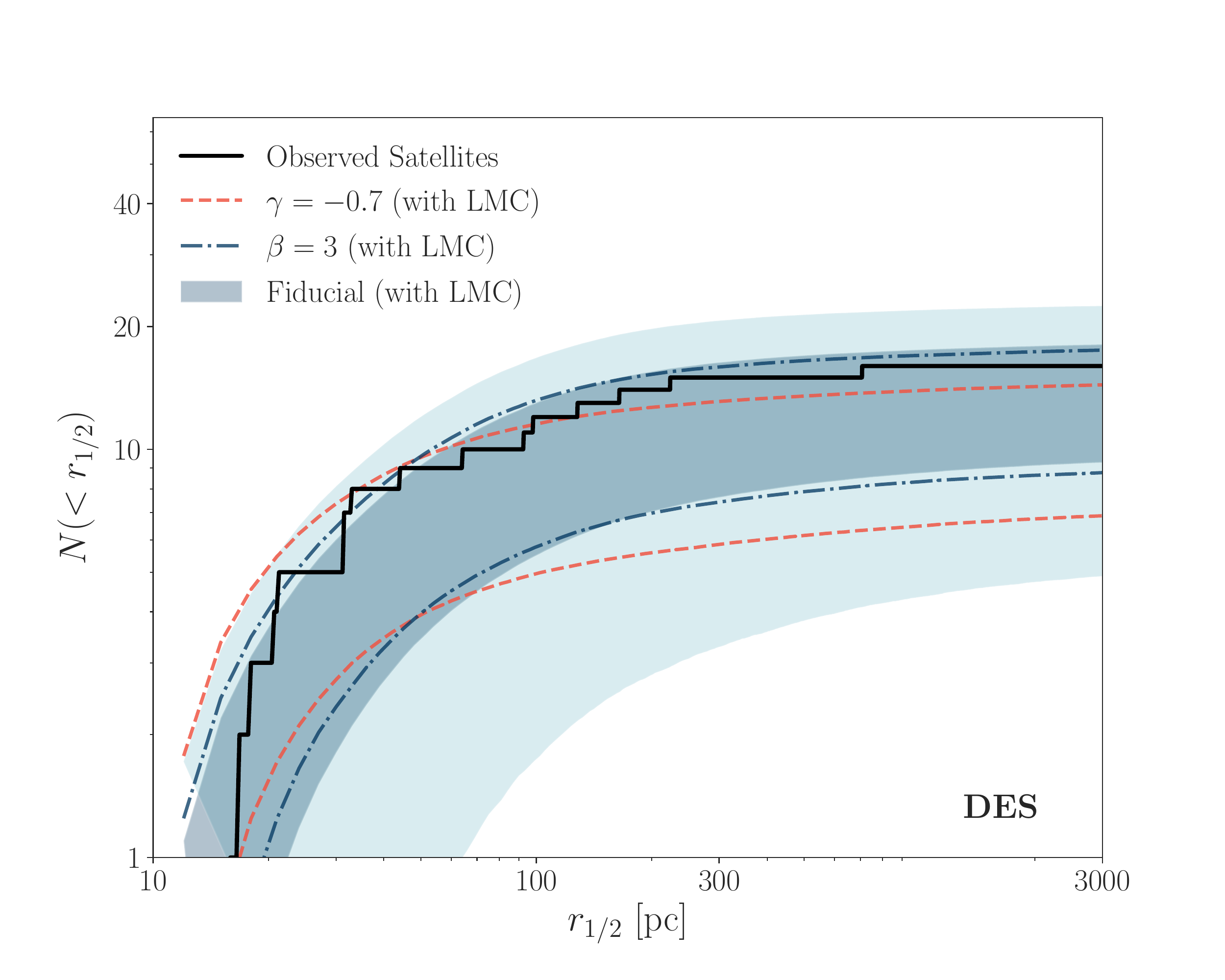}
    \includegraphics[scale=0.35]{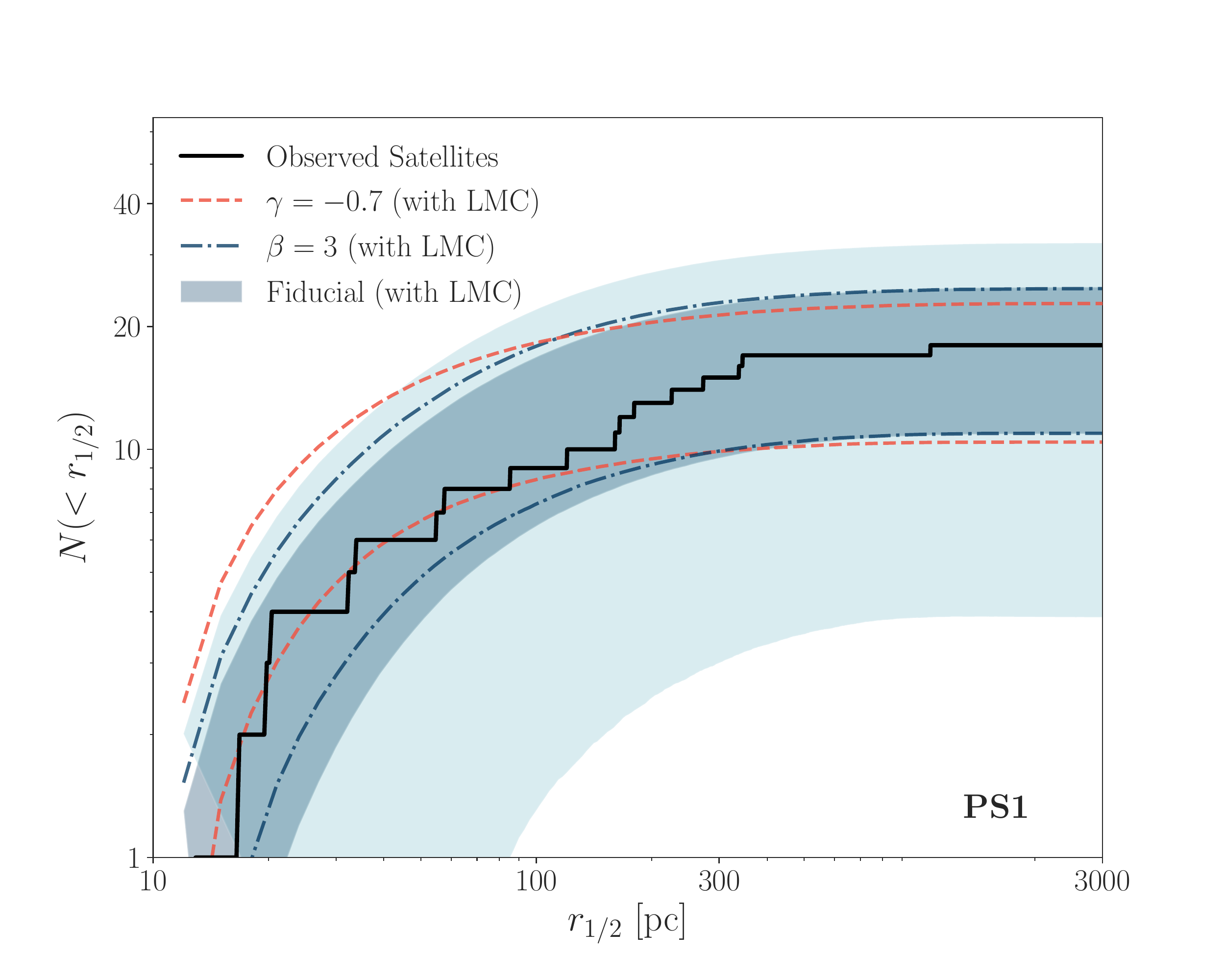}
    \caption{Size distributions derived by fitting to the DES and PS1 satellite populations. Our fiducial eight-parameter galaxy occupation fraction model is shown in blue. Dark (light) blue bands correspond to $68\%$ ($95\%$) confidence intervals, dashed red lines show the $68\%$ confidence interval for a model with a concentration-dependent galaxy--halo size relation, and dotted-dashed blue lines show the $68\%$ confidence interval for a model with an extreme dependence of satellite size on tidal stripping.}
    \label{fig:tidal_stripping}
\end{figure*}

To test whether a concentration-dependent size model is favored by the DES and PS1 data, we refit these satellite populations with $\gamma$ as an additional ninth free parameter. Because the concentration of subhalos after infall into the MW is difficult to measure accurately in our simulations, we measure the concentration at the time of accretion when implementing Equation \ref{eqn:size}. We find that our galaxy--halo connection model constraints are largely unchanged in this case, although the upper limits on $\sigma_{\log R}$ ($0.88\ \rm{dex}$) and $n$ (1.7) are more stringent than in our fiducial model. We find that the amplitude of the size relation is degenerate with $\gamma$, and our analysis does not place an upper limit on $\mathcal{A}$ in this case. Here $\gamma$ itself is constrained to lie between $-1.5$ and $-0.2$ at $95\%$ confidence. The predicted luminosity functions and size distributions are nearly identical to those from our fiducial analysis (we illustrate the size distribution for our fiducial posterior evaluated with $\gamma=-0.7$ in Figure \ref{fig:tidal_stripping}).

\subsection{Orphan Satellite Contribution}
\label{appendixa6}

To test the importance of orphan satellites, we refit the DES and PS1 satellite populations with $\mathcal{O}=0$, which adds zero orphans to our fiducial subhalo populations and effectively assumes that there is no artificial subhalo disruption in our simulations. Our constraints are virtually unaffected by this extreme variation in $\mathcal{O}$. In particular, the $95\%$ confidence upper limit on $\mathcal{M}_{50}$ increases by less than $1\sigma$ to $10^8\ M_{\rm{\odot}}$, and the rest of our galaxy--halo connection model constraints are also not significantly affected. The total number of MW satellites with $M_V < 0\ \rm{mag}$ and $r_{1/2}>10\ \rm{pc}$ decreases to~$190\pm 50$, as expected from the absence of an orphan satellite population. Thus, $\sim 15\%$ of the systems in our best-fit model are orphan satellites; these satellites might be associated with heavily stripped or disrupting subhalos.

\begin{deluxetable*}{{l@{\hspace{0.2in}}c@{\hspace{0.2in}}c}}[t]
\centering
\tablecolumns{3}
\tablecaption{Prior Distributions for the Parameters Varied in Our Fiducial Eight-parameter Fit to the DES and PS1 Satellite Populations}
\tablehead{\colhead{Free Parameter\phantom{texttttt}} {\hspace{0.2in}}& \colhead{Prior Distribution\phantom{text}} {\hspace{0.2in}}& \colhead{Motivation}}
\startdata
Faint-end Slope
& $\arctan\alpha\sim\rm{unif}(-1.1,-0.9)$ 
& Jeffreys prior for $-2<\alpha<-1.2$\\
\hline
Luminosity scatter
& $\sigma_M\sim\rm{unif}(0,2)\ \rm{dex}$
& Conservative upper limit \citep{GarrisonKimmel160304855,Lehmann151005651}\\
\hline
$50\%$ occupation mass
& $\log(\mathcal{M}_{50}/M_{\rm{\odot}})\sim\rm{unif}(7.5,11)$ 
& Lower limit corresponds to simulation resolution limit \citep{Mao150302637}\\
\hline
Baryonic effects
& $\ln(\mathcal{B})\sim \mathcal{N}(\mu=1,\sigma=0.5)$ 
& Hydrodynamic simulations \citep{Nadler171204467,Nadler180905542}\\
\hline
Occupation scatter
& $\sigma_{\rm{gal}}\sim\rm{unif}(0,1)\ \rm{dex}$ 
& Hydrodynamic simulations \citep{Fitts180106187,Graus180803654}\\
\hline
Size amplitude
& $\mathcal{A}\sim\rm{unif}(0,0.5)\ \rm{kpc}$ 
& Empirical galaxy--halo size relation \citep{Kravstov12122980}\\
\hline
Size scatter
& $\sigma_{\log R}\sim\rm{unif}(0,2)\ \rm{dex}$ 
& Empirical galaxy--halo size relation \citep{Kravstov12122980}\\
\hline
Size power-law index
& $n\sim \mathcal{N}(\mu=1,\sigma=0.5)$ 
& Empirical galaxy--halo size relation \citep{Kravstov12122980}\\
\hline
\enddata
{\footnotesize \tablecomments{Here $\mathcal{N}(\mu,\sigma)$ denotes a normal distribution with mean $\mu$ and standard deviation $\sigma$.}}
\label{tab:prior}
\end{deluxetable*}

\section{Statistical Framework Details}
\label{appendixb}

Next, we provide additional details on our statistical framework, and we discuss several caveats.

\subsection{Poisson Process Likelihood} 
\label{appendixb1}

The Poisson point process likelihood in our statistical comparison to observed satellites is implemented as follows. Suppose we observe $n_i$ real satellites and $\hat{n}_{i,\nu}$ mock satellites in an absolute magnitude bin $i$, where $\nu=1,\dots ,\hat{N}$ runs over all model realizations, including different host halos, and draws from our stochastic galaxy--halo connection model. The likelihood of observing these satellites given our model realizations, which enters Equation \ref{eq:likelihood}, is then
\begin{align}
P(n_{i}|\hat{n}_{i,1},\dots,\hat{n}_{i,\hat{N}}) &= \Big(\frac{\hat{N}+1}{\hat{N}}\Big)^{-(\hat{n}_{i,1} + \dots + \hat{n}_{i,\hat{N}} + 1)}&\nonumber\\ &\times (\hat{N}+1)^{-n_i}\frac{\Gamma(\hat{n}_{i,1} + \dots + \hat{n}_{i,\hat{N}} + n_i + 1)}{\Gamma(n_i + 1)\Gamma(\hat{n}_{i,1} + \dots + \hat{n}_{i,\hat{N}} + 1)},&\label{eq:like}
\end{align}
where the dependence on galaxy--halo connection model parameters $\boldsymbol{\theta}$ is implicit, and we assumed (i) a flat prior on $\lambda_i$ for $\lambda_i \geqslant 0$, and (ii) that $n_i$ and all $\hat{n}_{i,\nu}$ are drawn from the same Poisson distribution with rate parameter $\lambda_i$. Note that our method yields noninteger numbers of mock satellites by counting each system as $p_{\rm{detect}}\times (1-p_{\rm{disrupt}})\times f_{\rm{gal}}$ objects according to Equation \ref{eq:nsat}, so we have replaced factorials in the Poisson likelihood with appropriate gamma functions. Our results are unaffected if we enforce integer satellite counts by performing a binary mock observation of each predicted satellite according to its detection probability.

\subsection{Priors} 
\label{appendixb2}

We list the prior distributions used in our fiducial analysis in Table \ref{tab:prior}, several of which are informed by previous work. The prior on the faint-end slope is a noninformative Jeffreys prior \citep{Jethwa161207834}. The upper limit on the luminosity scatter is chosen to be very conservative; for example, \cite{Lehmann151005651} found that abundance-matching scatter at the luminosity scale of the brightest systems used in our analysis is less than $\sim 0.25\ \rm{dex}$. For $\mathcal{M}_{50}$, we set the lower limit of the prior based on the resolution limit of our simulations, which is a maximally conservative choice from the perspective of the inferred upper limit on this quantity. In particular, while we can decrease the lower limit of this prior because the $\mathcal{M}_{50}$ posterior is flat below~$\sim 5\times 10^{7}\ M_{\rm{\odot}}$ due to the limited sensitivity of the DES and PS1 satellite searches, doing so would artificially decrease the inferred $95\%$ confidence upper limit.\footnote{However, as noted in \S\ref{constraints}, our reported Bayes factors are independent of this choice.} Priors for $\mathcal{B}$ and $n$ are set based on studies that identify the preferred values of these parameters, and priors for $\sigma_{\rm{gal}}$, $\mathcal{A}$, and $\sigma_{\log R}$ are chosen to be uniform with conservative upper bounds.

\subsection{Bayes Factor Calculation} 
\label{appendixb3}

To calculate Bayes factors, we estimate the Bayesian evidence using the bounded harmonic mean method described in \cite{Nadler180905542}. In particular, for a given posterior, we select samples of galaxy--halo connection model parameters $\boldsymbol{\theta}$ within a fixed Mahalanobis distance of a point $\boldsymbol{\theta}_0$ in a high-density region of the posterior. We then average the inverse of the posterior probabilities for these samples, and we normalize by the volume of the sampled region. We repeat this procedure for high-density regions that contain~$10\%$--$25\%$ of the total number of MCMC samples, and we average over these percentiles to obtain the mean Bayesian evidence.

\subsection{Caveats and Future Work} 
\label{appendixb4}

In this work, we fit to observed MW satellites in an observable parameter space $x$ that consists of absolute magnitude and two large surface brightness bins. However, it would be more constraining to perform our inference in a higher-dimensional space that includes galactocentric distance. There are two main difficulties inherent in our statistical modeling.
\begin{enumerate}
\item We have binned observed and modeled satellites assuming that the unknown Poisson process rate in each bin is independent from the rate in other bins. This assumption is unphysical, as the rate should vary smoothly in observable parameter space.
\item As the number of bins increases, the number of satellites per bin decreases, which causes the uncertainty in the rate parameter to increase and our model to become increasingly unconstrained. This is a particularly challenging problem as we move to higher-dimensional parameter spaces, since the number of bins increases rapidly with dimensionality.
\end{enumerate}
To address these issues, it is possible to connect rates in nearby regions of parameter space in an unbinned fashion using a correlated prior. This is equivalent to imposing that our galaxy--halo model should produce satellite abundances that vary smoothly as a function of observable quantities. We now lay out the mathematical formalism necessary for introducing this prior.

Our model of the distribution of satellites in observable space is an inhomogeneous Poisson process, where the number of ``events'' in any region $\mathcal{T}$ of observable space $x$ is given by a Poisson distribution with rate $\lambda_{\mathcal{T}} = \int_{\mathcal{T}} \lambda \left( x \right) \mathrm{d}x$, where~$\lambda \left(x\right)$ is referred to as the ``rate function.'' Given a rate function $\lambda \left( x \right)$, the likelihood of observing $N$ events at a set of points $\left\{ x_i \right\}_{i=1}^N$ is
\begin{equation}
      p \left( \left\{ x_i \right\}_{i=1}^N \mid \lambda \right)
  =
  \exp \left[ -\int \lambda \left( x \right) \mathrm{d}x \right]
  \prod_{i=1}^N \lambda \left( x_i \right),
\end{equation}
where we suppressed the dependence of the rate on our galaxy--halo connection model parameters $\boldsymbol{\theta}$. In our case, the ``events'' $\left\{ x_i \right\}_{i=1}^N$ are the locations of detected satellites in an observable parameter space. Note that in this formulation, there is no binning in $x$.

Calculating this likelihood exactly is challenging because, in order to compare observed and modeled satellite populations, we must integrate over the unknown rate function $\lambda$,
\begin{equation}
  p \left( \left\{ x_i \right\} \mid \left\{ \hat{x}_j \right\} \right)
  =
  \frac{
    \int \mathcal{D} \lambda \,\,
    p \left( \left\{ x_i \right\} \mid \lambda \right)
    p \left( \left\{ \hat{x}_j \right\} \mid \lambda \right)
    p \left( \lambda \right)
  }{
    \int \mathcal{D} \lambda \,\,
    p \left( \left\{ \hat{x}_j \right\} \mid \lambda \right)
    p \left( \lambda \right)
  }
  \, .\label{eq:functional}
\end{equation}
Here, both the numerator and denominator contain functional integrals over the rate; these integrals are performed over an infinite-dimensional space consisting of the rate at each point in observable parameter space. Further, this rate is a stochastic function in our galaxy--halo connection model due to satellite luminosity and size scatter. This makes our model an inhomogeneous Poisson process with a stochastic rate function, which is known as a ``Cox process.'' The prior on the rate function, $p \left( \lambda \right)$, must admit only positive rates; one possible choice is to treat the logarithm of the rate as a Gaussian process. Models involving Cox processes are often termed ``doubly intractable'' due to the presence of intractable integrals over the rate function \citep{Murray2007PhD}.

There are, however, several approaches to make Cox processes tractable. As noted above, we bin satellites in absolute magnitude and split the sample into two large surface brightness bins, so that our likelihood is over the number of counts in each bin, rather than the locations of the points. This is equivalent to assuming that the rate function is constant in each bin and leads to the likelihood
\begin{equation}
  p \left( \left\{ n_j \right\}_{j=1}^{N_{\mathrm{bins}}} \mid \left\{ \lambda_j \right\} \right)
  =
  \exp \left( -\sum_{\mathrm{bins}\ j} \lambda_j \mathcal{V}_j \right)
  \prod_{\mathrm{bins}\ k} \frac{\lambda_k^{n_k}}{n_k !}
  \, ,
\end{equation}
where $\lambda_j$ is the rate in bin $j$, $\mathcal{V}_j$ is the volume of bin $j$, and $n_j$ is the number of events in bin $j$. Binning turns the functional integral over $\lambda \left( x \right)$ in Equation \ref{eq:functional} into a finite-dimensional integral over the value of $\lambda$ in each bin. Choosing Cartesian bins in observable parameter space then renders the problem tractable \citep{Flaxman}. There also exist approaches that avoid binning the observable space \citep{Adams,John180401016}, which we intend to explore in future work.

\begin{deluxetable}{l @{\hspace{0.485in}}c@{\hspace{0.485in}}c@{\hspace{0.485in}}c}[t]
\tabletypesize{\footnotesize}
\tablecolumns{4}
\tablecaption{MW Satellites Used in Our Analysis.}
\tablehead{
\colhead{Name\phantom{texttttt}}{\hspace{0.485in}} & \colhead{$M_V$}{\hspace{0.485in}} & \colhead{$D$}{\hspace{0.485in}} & \colhead{$r_{1/2}$} \\
  \colhead{\phantom{Name}}{\hspace{0.5in}} & \colhead{(mag)}{\hspace{0.485in}} & \colhead{(kpc)}{\hspace{0.485in}} & \colhead{(pc)}}
\startdata
& \small{DES}  & & \\
Fornax & -13.46 & 147 & 707\\
Sculptor & -10.82 & 84 & 223\\
Reticulum II & -3.88 & 30 & 31\\
Eridanus II\tablenotemark{\footnotesize{a}} & -7.21 & 380 & 158\\
Tucana II & -3.8 & 58 & 165\\
Grus II$^*$ & -3.9 & 53 & 92\\
Horologium I & -3.55 & 79 & 31\\
Tucana III$^*$ & -2.4 & 25 & 44\\
Tucana IV & -3.5 & 48 & 128\\
Phoenix II & -3.30 & 83 & 21\\
Horologium II$^*$ & -2.6 & 78 & 33\\
Tucana V$^*$ & -1.6 & 55 & 16\\
Pictor I$^*$ & -3.45 & 114 & 18\\
Columba I$^*$ & -4.2 & 183 & 98\\
Cetus II$^*$ & 0.02 & 30 & 17\\
Grus I$^*$ & -3.47 & 120 & 21\\
Reticulum III$^*$ & -3.31 & 92 & 64\\
\hline
& \small{PS1}  & & \\
Leo I & -11.78 & 254 & 226\\
Leo II & -9.74 & 233 & 165\\
Draco & -8.71 & 76 & 180\\
Ursa Minor & -9.03 & 76 & 272\\
Sextans & -8.72 & 86 & 345\\
Canes Venatici I & -8.8 & 218 & 338\\
Bo\"otes I & -6.02 & 66 & 160\\
Ursa Major II & -4.25 & 32 & 85\\
Coma Berenices & -4.38 & 44 & 57\\
Sagittarius II & -5.2 & 69 & 32\\
Willman 1 & -2.53 & 38 & 20\\
Canes Venatici II & -5.17 & 160 & 55\\
Segue 1 & -1.30 & 23 & 20\\
Segue 2$^*$ & -1.86 & 35 & 34\\
Crater II & -8.2 & 117 & 1066\\
Draco II$^*$ & -0.8 & 22 & 17\\
Triangulum II$^*$ & -1.60 & 30 & 13\\
Hercules & -5.83 & 132 & 120\\
Cetus II\tablenotemark{\footnotesize{b}} & 0.02 & 30 & 17\\
\enddata
{\footnotesize \tablecomments{Properties of confirmed and candidate DES and PS1 satellites used in our analysis, listed in order of detection significance (\citetalias{PaperI}). Asterisks mark kinematically unconfirmed systems.}}
\tablenotetext{a}{\footnotesize{Eridanus II is not included because it lies outside our fiducial $300\ \rm{kpc}$ heliocentric distance cut.}}
\tablenotetext{b}{\footnotesize{Cetus II is detected in both PS1 and DES; in our analysis, we only count this system in the observed DES population.}}
\label{fig:des_table}
\end{deluxetable}

\section{Robustness to Observational Systematics}
\label{appendixd}

We now present a set of tests in order to verify the robustness of our key results to various observational systematics.

\subsection{Kinematically Unconfirmed Satellites} 
\label{appendixd1}

\begin{figure*}[ht]
\centering
    \includegraphics[scale=0.61]{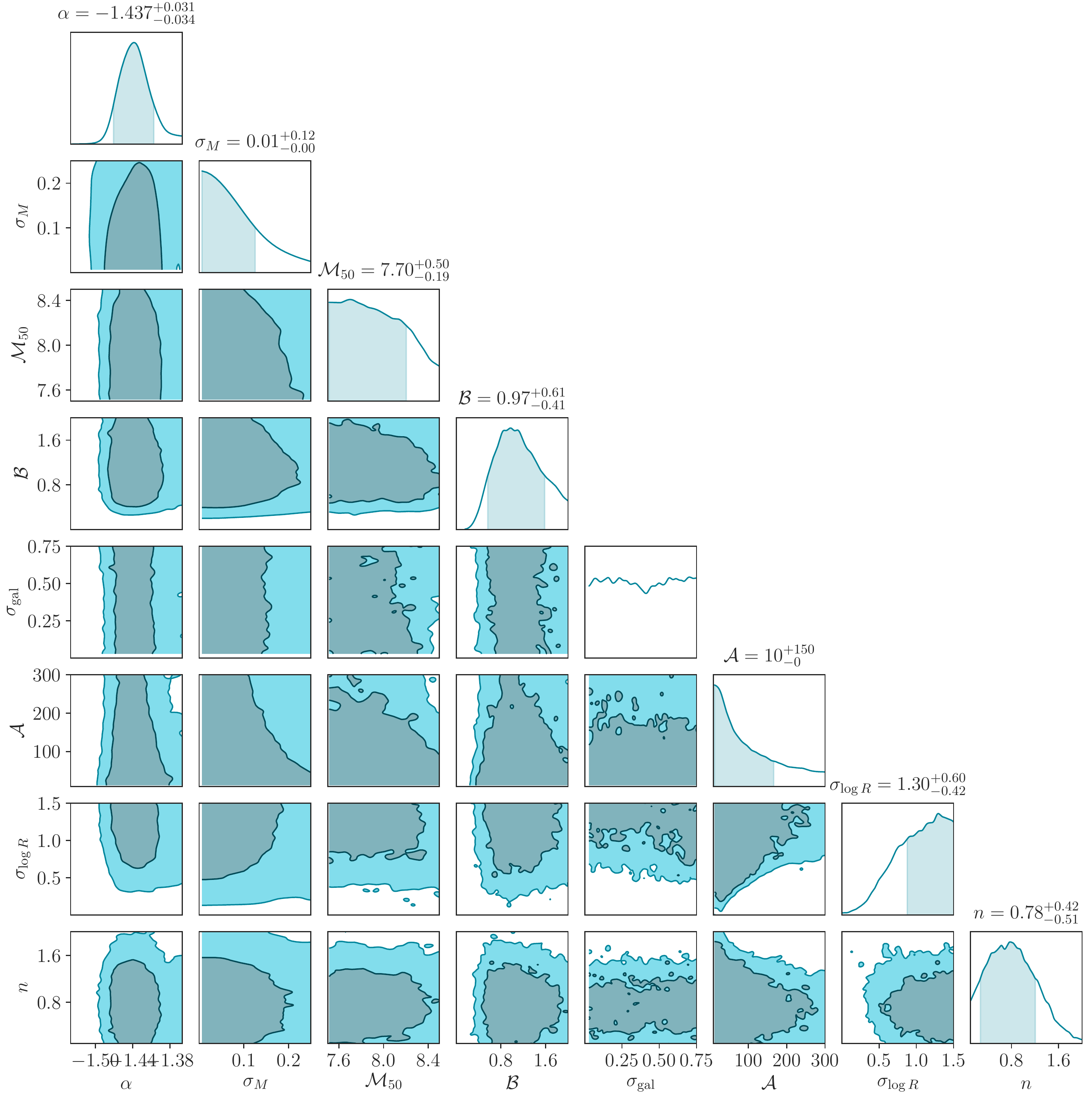}
    \caption{Posterior distribution from our fit to the kinematically confirmed DES and PS1 satellite populations. Dark (light) shaded contours represent $68\%$ ($95\%$) confidence intervals. Shaded areas in the marginal distributions and parameter summaries correspond to $68\%$ confidence intervals. Note that $\sigma_M$, $\sigma_{\rm{gal}}$, and $\sigma_{\log R}$ are reported in $\rm{dex}$, $\mathcal{M}_{50}$ is reported as $\log(\mathcal{M}_{50}/M_{\rm{\odot}})$, and $\mathcal{A}$ is reported in $\rm{pc}$. Note that $\sigma_{\rm{gal}}$ is not constrained at $68\%$ confidence in this fit.}
    \label{fig:posterior_conservative}
\end{figure*}

To assess possible systematic uncertainties associated with the observed set of DES and PS1 satellites presented in \citetalias{PaperI}, we rerun the entire analysis using only satellites that have are confirmed to exhibit dark matter--dominated internal kinematics. The candidate satellites excluded from this reanalysis are indicated in Table \ref{fig:des_table}. As shown in Figure \ref{fig:posterior_conservative}, our galaxy--halo connection model constraints are largely unaffected by refitting the DES and PS1 satellite populations under the conservative assumption that all unconfirmed systems are star clusters. Most importantly, the upper limit on $\mathcal{M}_{50}$ only increases by $\sim 1\sigma$, to $5\times 10^{8}\ M_{\rm{\odot}}$ at $95\%$ confidence, and the minimum halo mass increases to $6.5\times 10^8\ M_{\rm{\odot}}$, similar to the minimum halo mass inferred from classical and SDSS satellites in \cite{Nadler180905542}. In addition, the total predicted number of MW satellites decreases by~$\sim 1\sigma$ to~$150 \pm 60$. These shifts are expected, since unconfirmed satellite candidates constitute many of the faintest systems in our fiducial sample. Thus, we conclude that our key constraints and predictions are not highly sensitive to the nature of kinematically unconfirmed satellite candidates.

\subsection{Satellite Size Criterion}
\label{appendixd2}

Next, we test whether a more conservative satellite size criterion impacts our results. For this test, we self-consistently exclude all observed and predicted satellites with $r_{1/2}>20\ \rm{pc}$ from our statistical inference, rather than the $r_{1/2}>10\ \rm{pc}$ cut used in our fiducial analysis. Our key constraints are not significantly affected; for example, the $95\%$ confidence level upper limit on $\mathcal{M}_{50}$ increases slightly, to $1.5\times 10^{8}\ M_{\rm{\odot}}$. The upper limit on the amplitude of the galaxy--halo size relation, which was $110\ \rm{pc}$ in our fiducial analysis, increases to~$220\ \rm{pc}$, as we might expect from excluding small satellites in the fit.

\subsection{Biases in Measured Satellite Properties}
\label{appendixd3}

Finally, we test whether systematic offsets in measured satellite properties could affect our conclusions. In particular, we assume that every measured DES and PS1 satellite absolute magnitude is offset from the fiducial value listed in Table \ref{fig:des_table} by $\Delta M_V = +1\ \rm{mag}$, which is similar to the width of the absolute magnitude bins used in our fiducial analysis. We rerun the entire analysis with these shifted magnitudes, and we repeat this procedure for $\Delta M_V = -1\ \rm{mag}$. In both cases, we still obtain a good joint fit to the DES and PS1 luminosity functions. As expected, the inferred faint-end slope is steeper (shallower) than that obtained from our fiducial analysis for $\Delta M_V = -1\ \rm{mag}$ ($\Delta M_V = +1\ \rm{mag}$); however, the total predicted number of MW satellites with $M_V < 0\ \rm{mag}$ and $r_{1/2}>10\ \rm{pc}$ and our $95\%$ confidence upper limit on $\mathcal{M}_{50}$ are not significantly affected in either case.

\section{Resolution and Sample Variance} 
\label{appendixd4}

To assess the impact of resolution effects on our fiducial simulations and results, we compare the subhalo maximum circular velocity function, radial distribution, and size distribution from one of our fiducial host halos (excluding LMC satellites) to those from a higher-resolution resimulation of the same host. In particular, we resimulate this halo with a $4\times 10^{4}\ M_{\mathrm \odot}\ h^{-1}$ high-resolution particle mass and an~$85\ \mathrm{pc}\ h^{-1}$ minimum softening length. We find that the distributions of all relevant subhalo properties are not significantly affected above the resolution limit of our fiducial simulations. Moreover, by rerunning our analysis, we find that none of our galaxy--halo connection model constraints are significantly affected when using a higher-resolution simulation.

We also assess the impact of sample variance on our fiducial subhalo and satellite populations, since the final positions of LMC satellites might be sensitive to the realizations of small-scale density fluctuations in our fiducial simulations. In particular, we resimulate both of our fiducial host halos at standard resolution with different random seeds for small-scale phases in the matter power spectrum below $60\ \mathrm{kpc}\ h^{-1}$. We find that the properties of the MW host halo and LMC halo are not significantly affected in these resimulations, and that the resulting subhalo populations are nearly identical in terms of their distributions of $\mathcal{M}_{\rm{peak}}$, $V_{\rm{peak}}$, halo size at accretion, and present-day heliocentric distance, implying that our results are robust to sample variance in the phases of the matter power spectrum on small scales.

\section{Observed Satellite Data Vectors}
\label{appendixe}

The confirmed and candidate DES and PS1 satellites that pass the detection criteria defined in \citetalias{PaperI} are listed in Table~\ref{fig:des_table}. Note that, although Kim 2 (DES) and Laevens 1 (PS1) formally pass these detection criteria, we do not include these systems in our analysis or Table \ref{fig:des_table} because they are suspected to be star clusters (\citetalias{PaperI}).

\end{document}